\documentclass[12pt,english]{article}
\usepackage{babel}
\usepackage{graphicx}
\usepackage{color}                    
\usepackage{multirow}
\begin{document}
\baselineskip 18pt
\def\today{\ifcase\month\or
 January\or February\or March\or April\or May\or June\or
 July\or August\or September\or October\or November\or December\fi
 \space\number\day, \number\year}

%
\def\thebibliography#1{\section*{References\markboth
 {References}{References}}\list
 {[\arabic{enumi}]}{\settowidth\labelwidth{[#1]}
 \leftmargin\labelwidth
 \advance\leftmargin\labelsep
 \usecounter{enumi}}
 \def\newblock{\hskip .11em plus .33em minus .07em}
 \sloppy
 \sfcode`\.=1000\relax}
\let\endthebibliography=\endlist
\def\beq{\begin{equation}}
\def\eeq{\end{equation}}
\def\beqn{\begin{eqnarray}}
\def\eeqn{\end{eqnarray}}
\def\rmuu{\gamma^{\mu}}
\def\rmud{\gamma_{\mu}}
\def\PL{{1-\gamma_5\over 2}}
\def\PR{{1+\gamma_5\over 2}}
\def\sinW2{\sin^2\theta_W}
\def\AEM{\alpha_{EM}}
\def\mul{M_{\tilde{u} L}^2}
\def\mur{M_{\tilde{u} R}^2}
\def\mdl{M_{\tilde{d} L}^2}
\def\mdr{M_{\tilde{d} R}^2}
\def\mz2{M_{z}^2}
\def\c2b{\cos 2\beta}
\def\au{A_u}
\def\ad{A_d}
\def\cob{\cot \beta}
\def\v#1{v_#1}
\def\tb{\tan\beta}
\def\epem{$e^+e^-$}
\def\KK{$K^0$-$\overline{K^0}$}
\def\wi{\omega_i}
\def\xj{\chi_j}
\def\Wmu{W_\mu}
\def\Wnu{W_\nu}
\def\m#1{{\tilde m}_#1}
\def\mH{m_H}
\def\mw#1{{\tilde m}_{\omega #1}}
\def\mx#1{{\tilde m}_{\chi^{0}_#1}}
\def\mc#1{{\tilde m}_{\chi^{+}_#1}}
\def\mwi{{\tilde m}_{\omega i}}
\def\mxi{{\tilde m}_{\chi^{0}_i}}
\def\mci{{\tilde m}_{\chi^{+}_i}}
\def\mz{M_z}
\def\sw{\sin\theta_W}
\def\cw{\cos\theta_W}
\def\cb{\cos\beta}
\def\sb{\sin\beta}
\def\rwi{r_{\omega i}}
\def\rxj{r_{\chi j}}
\def\rfp{r_f'}
\def\Kik{K_{ik}}
\def\Fq2{F_{2}(q^2)}
\def\f{\({\cal F}\)}
\def\d1{{\f(\tilde c;\tilde s;\tilde W)+ \f(\tilde c;\tilde \mu;\tilde W)}}
\def\tw{\tan\theta_W}
\def\sec2w{sec^2\theta_W}

\def\bl{$B\&L$}

\begin{titlepage}

\begin{center}
{\Huge {\ { \textsf{
Yukawa Couplings and
Quark and Lepton Masses
   in an $SO(10)$  Model with a Unified Higgs Sector
    }}}}\\

\vskip 0.5 true cm \vspace{2cm}
\renewcommand{\thefootnote}
{\fnsymbol{footnote}}
 Pran Nath$^a$ and Raza M. Syed$^{b}$
\vskip 0.5 true cm
\end{center}
\begin{center}
\noindent {a. Department of Physics, Northeastern University,
Boston, MA 02115-5000, USA \\
b. Department of Physics, American University of Sharjah,
P.O. Box 26666,
Sharjah, UAE
} \\
\end{center}
\vskip 1.0 true cm \centerline{\bf Abstract}
An analytic computation is  given of the generation of Yukawa couplings and of the
quark, charged lepton and neutrino masses
in the framework of an $SO(10)$ model with  a unified Higgs sector consisting of a  single pair of
vector-spinor $144 + \overline{144}$ of Higgs multiplets. This unified Higgs sector allows
for a breaking of $SO(10)$ to the gauge group $SU(3)\times SU(2)_L \times U(1)_Y$
and contains light Higgs doublets allowing for the breaking of the electroweak
symmetry. Fermion mass generation in this model typically arises from quartic
couplings
 $16\cdot 16\cdot 144\cdot  144$ and
$16\cdot 16\cdot \overline{144}\cdot  \overline{144}$.
Extending a previous work it is shown
that much larger third generation
masses can arise for all the  fermions from mixing with $45$ and $120$ matter  multiplets
via the cubic couplings $16\cdot45\cdot\overline{144}$ and $16\cdot120\cdot144$.
Further it is  found that values of $\tan\beta$ as low as 10 can allow for a $b-\tau-t$
unification consistent with current data. The quartic and cubic couplings naturally lead to Dirac  as well as
Majorana neutrino masses necessary for the  generation of See Saw neutrino
masses.

\medskip
\noindent

\end{titlepage}

\newcommand{\Nu}{\mbox{\LARGE$\nu$}}
\newcommand{\Tau}{\mbox{\LARGE$\tau$}}
\newcommand{\bNu}{\mbox{\boldmath$\nu$}}
\newcommand{\bTau}{\mbox{\boldmath$\tau$}}
\newcommand{\mTau}{\mbox{\normalsize$\tau$}}
\newcommand{\mNu}{\mbox{\normalsize$\nu$}}

\section{Introduction\label{1}}
Over the recent past the gauge group $SO(10)$ has emerged as the  leading candidate for
the unification of  forces including the strong, the weak and the electromagnetic\cite{georgi},
and at the same time the $16$ plet representation of $SO(10)$ can accommodate a full
generation of  quarks and leptons. Further, the singlet in the $16$ plet is identified
as a right handed neutrino which can combine with the left handed  neutrino in the $16$ plet
to generate a Dirac mass and at the same time if the singlet acquires a large Majorana mass, one can
generate small neutrino masses by the well known See Saw mechanism. While this is a very nice
picture, an actual implementation of an
 $SO(10)$ grand unification requires  many Higgs multiplets
for the  explicit breaking of the gauge symmetry down to its residual $SU(3)_C\times U(1)_{\rm em}$
form (For a review  see \cite{Nath:2006ut}).
Thus, for example, a possible breaking scheme discussed in the literature is to use
a $45$ plet of Higgs to break $SO(10)$,
 a $16+\overline {16}$ to reduce the rank of the gauge group and  $10$ plets of Higgs to
 break the electroweak symmetry. Many other schemes involving several Higgs multiplets for the breaking of $SO(10)$ to its residual $SU(3)_C\times U(1)_{\rm em}$ have also been discussed. \\

 Recently a new formalism was proposed which has a unified Higgs structure, i.e., a
 $144+\overline{144}$ multiplets of  Higgs \cite{Babu:2005gx,Babu:2006rp}.
  It was shown that the above Higgs structure
 can break $SO(10)$ and and reduce the rank of the gauge group at the same time
 such that
   $SO(10)\to SU(3)_C\times SU(2)\times U(1)_Y$ at the GUT scale $M_G$. How this comes
about can be easily seen by examining the $SU(5)\times U(1)$ decomposition of the
$45$  plet and the $144$ plet in representations of $SO(10)$   so that
\beqn\label{1.1}
45= 1(0) + 10(4) +\overline{10}(-4) +24(0),\nonumber\\
144= \bar 5(3) + 5(7) + 10(-1)+{15}(-1) + 24(-5) +40(-1) +\overline{45}(3).
\eeqn
where the quantity in the parentheses represents the $U(1)$ quantum numbers.
For the usual model building where the $45$ plet of Higgs is utilized to break $SO(10)$ it is
a combination of the singlet $1(0)$ and the $24(0)$ plets of Higgs that develop VEVs. Since
each of these carry no $U(1)$ quantum numbers, giving a VEV to the $45$ plet does not
reduce the rank of the group. In the decomposition of $144$ one finds that all the components
carry $U(1)$ quantum numbers, and thus a VEV formation in $144$ along the $45$ plet direction
not only breaks  $SO(10)$ but
also reduces the rank of the gauge group and consequently $SO(10)$ breaks directly to the
Standard Model gauge group $SU(3)_C \times SU(2)_L\times U(1)_Y$.
Further, it   was shown in \cite{Babu:2005gx,Babu:2006rp}
that a pair of light doublets can be gotten which are necessary for the breaking
of the electroweak symmetry which occurs when $5, \bar 5, \overline{45} (45)$ components of
$144(\overline{144})$ develop vacuum expectation values.
 With the $144+\overline{144}$ Higgs sector, mass generation for the fermions
requires at least  the quartic interactions
\beqn\label{1.2}
\frac{1}{\Lambda}(16\cdot16\cdot144\cdot144), ~~\frac{1}{\Lambda}(16\cdot16\cdot\overline{144}\cdot\overline{144}).
\eeqn
where $\Lambda$ would typically be of size  the string scale  $M_{st}$.
 After spontaneous breaking at
the GUT scale  the $144$ and $\overline{144}$ will develop a VEV of size $M_G$ and the above quartic coupling will lead to
 Yukawa interactions of matter fields with the Higgs doublets and these Yukawas
    will be  $O(M_G/M_{st})$, which  are the appropriate  size for the Yukawas for the
first and for the second generation fermions. One of the aims of the analysis of this work  is to compute
explicitly these Yukawa couplings.
To exhibit this explicitly
we expand the $16$ plet of matter fields in $SU(5)\times U(1)$ decomposition so that
\beqn\label{1.3}
16=1(-5) + \bar 5(3) + 10(-1).
\eeqn
It is then easily seen that interactions of Eq.(\ref{1.2}) produce the Yukawa couplings necessary for
generating the masses for  the
up quarks, the down quarks and the charged leptons, and also give couplings which generate
 Dirac and Majorana masses for the neutrinos. We list these below
in the $SU(5)\times U(1)$ decomposition indicating the light doublet ${\mathsf D}_1$, ${\mathsf D}_2$ or ${\mathsf D}_3$ for
which the specific interaction listed below is valid\footnote{The doublet ${\mathsf D_1}$ arises from the $5$ and $\bar 5$ plets such that
${\mathsf D}_1 \subset (5_{144}(7),\bar 5_{\overline{144}}(-7))$.
Similarly ${\mathsf D_2}$ and ${\mathsf D_3}$ are linear combination of doublets arising from  the following
$(\bar 5_{144}(3), 5_{\overline{144}}(-3))$ and $(\overline{45}_{144}(3), {45}_{\overline{144}}(-3))$.
More precise definitions of ${\mathsf D}_1$, ${\mathsf D}_2$ and ${\mathsf D}_3$ are given in Eq.(\ref{3.7}) and Eq.(\ref{3.8}).
}
\beqn
\underline{{\rm From~ quartic ~couplings } ~16\cdot16\cdot144\cdot144}&&\nonumber\\
{u - \rm quarks}:&& \frac{1}{\Lambda}10_{16}(-1)\cdot10_{16}(-1)\cdot<5_{144}(7)\cdot24_{144}(-5)> ~({\mathsf D}_1),\label{1.41}\\
{d - \rm quarks +leptons}:&& \frac{1}{\Lambda}\bar 5_{16}(3)\cdot10_{16}(-1)\cdot<\bar 5_{144}(3)\cdot24_{144}(-5)>
~({\mathsf D}_2, {\mathsf D}_3),\label{1.42}\\
&&\frac{1}{\Lambda}\bar 5_{16}(3)\cdot10_{16}(-1)\cdot<{\overline{45}}_{144}(3)\cdot24_{144}(-5)> ~({\mathsf D}_2,{\mathsf D}_3),\label{1.421}\\
{LR}-\nu{\rm ~mass}: && \frac{1}{\Lambda} 1_{16}(-5)\cdot\bar 5_{16}(3)\cdot<5_{144}(7)\cdot24_{144}(-5)>~({\mathsf D}_1),\label{1.43}\\
{RR}-\nu{\rm ~mass}: && {\rm None},\label{1.44}\\
{LL}-\nu{\rm ~mass}: && \bar 5_{16}(3)\cdot\bar 5_{16}(3)\cdot15_{144}(-1)\cdot24_{\overline{144}}(-5)~({\mathsf D}_1, {\mathsf D}_2, {\mathsf D}_3),\label{1.45}
\eeqn
where $10_{16}(-1)$ etc. stand for the  $10$ plet of $SU(5)$ arising from the $16$ plet of $SO(10)$ and $(-1)$ is
the $U(1)$ charge in the $SO(10)\to SU(5)\times U(1)$ decomposition etc..
\beqn
\underline{{\rm From~ quartic ~couplings} ~16\cdot 16\cdot\overline {144}\cdot\overline {144}}&&\nonumber\\
{ u - \rm quarks:}  &&\frac{1}{\Lambda}10_{16}(-1)\cdot10_{16}(-1)\cdot<5_{\overline {144}}(-3)\cdot24_{\overline{144}}(5)> ~({\mathsf D}_2,{\mathsf D}_3),\label{1.51}\\
&&\frac{1}{\Lambda}10_{16}(-1)\cdot10_{16}(-1)\cdot<{45}_{\overline {144}}(-3)\cdot24_{\overline{144}}(5)> ~({\mathsf D}_2,{\mathsf D}_3),\label{1.51}\\
 { d - \rm quarks +leptons}:  &&\frac{1}{\Lambda}\bar 5_{16}(3)\cdot10_{16}(-1)\cdot<\bar 5_{\overline{144}}(-7)\cdot24_{\overline{144}}(5)>~({\mathsf D}_1),\label{1.52}\\
{LR}-\nu{\rm ~mass}:  &&\frac{1}{\Lambda} 1_{16}(-5)\cdot\bar 5_{16}(3)\cdot<5_{\overline{144}}(-3)\cdot24_{\overline{144}}(5)>~({\mathsf D}_2,{\mathsf D}_3),\label{1.53a}\\
&&\frac{1}{\Lambda} 1_{16}(-5)\cdot\bar 5_{16}(3)\cdot<{45}_{\overline{144}}(-3)\cdot24_{\overline{144}}(5)>~
({\mathsf D}_2,{\mathsf D}_3),\label{1.53}\\
{RR}-\nu{\rm ~mass}:  &&\frac{1}{\Lambda}1_{16}(-5)\cdot1_{16}(-5)\cdot<24_{\overline{144}}(5)\cdot24_{\overline{144}}(5)>~({\mathsf D}_1, {\mathsf D}_2, {\mathsf D}_3),\label{1.54}\\
{LL}-\nu{\rm ~mass}:  &&\frac{1}{\Lambda}\bar 5_{16}(3)\cdot\bar 5_{16}(3)\cdot<5_{\overline{144}}(-3)\cdot 5_{\overline{144}}(-3)>~({\mathsf D}_2,{\mathsf D}_3),\label{1.55}\\
&&\frac{1}{\Lambda}\bar 5_{16}(3)\cdot\bar 5_{16}(3)\cdot<{45}_{\overline{144}}(-3)\cdot {45}_{\overline{144}}(-3)>~({\mathsf D}_2,{\mathsf D}_3).\label{1.55}
\eeqn
After spontaneous breaking at the GUT scale the Higgs multiplets $24_H$ develop a VEV of size $M_G$ generating Yukawa couplings for the quarks and the
leptons of size $O(M_G/\Lambda)$ followed by a spontaneous symmetry breaking at the electroweak
scale which  give VEVs to the Higgs doublets and
generate quark and lepton masses.

Since $\Lambda$ would  be of the size of the string scale
($M_{St}$) the ratio $O(M_G/\Lambda)<<1$ is appropriate for the Yukawa couplings for the first
two generations of quarks and charged leptons. From Eq.(\ref{1.53a}) and
Eq.(\ref{1.53}) one finds that a similar
size $LR$ Yukawa coupling arises for the neutrinos. Quite remarkably the same quartic interactions of
 Eq.(\ref{1.2})  lead to large (GUT size) masses needed for the generation of Type I See Saw
 neutrino masses and very small  $LL$ neutrino masses for a Type II See Saw.
 Thus from Eq.(\ref{1.54}) we see that with $M_G\sim 10^{16}$ GeV, and $\Lambda \sim 10^{17}$ GeV
 one finds that $RR$ neutrino masses are $O(M_G^2/\Lambda)\sim 10^{15}$ GeV.
 Similarly from Eq.(\ref{1.55}) one finds the $LL$ neutrino masses to be of size $O(M_{EW}^2/\Lambda)$ which gives
 $LL$ masses in the range $10^{-3\pm 1}$eV .\\

\noindent
We note now the following two possibilities:
\begin{enumerate}
\item For  the case when
$(\bar 5_{\overline{144}}(-3),5_{144}(3))$ and  $(45_{\overline{144}}(-3), \overline{45}_{144}(3))$
contain
 the light pair of doublets,  only the down quark masses will arise from the quartic couplings
$16\cdot16\cdot144\cdot144$, while the  masses for the up quarks,  $LR$-Dirac masses, $LL$ and
$RR$ masses for the neutrinos will all arise only from the   $16\cdot16\cdot\overline {144}\cdot\overline {144}$ couplings. \\
\item For  the case when
$(\bar 5_{\overline {144}}(-7),5_{144}(7))$  contain the light doublets,
  the up quark masses and the $LR$-Dirac neutrino masses
arise from the quartic couplings $16\cdot16\cdot144\cdot144$, while the  masses
for the down quark and lepton  as well as $RR$ neutrino masses
  arise from the   $16\cdot16\cdot\overline {144}\cdot\overline {144}$ couplings. Here, however,  no $LL$
   neutrino masses are generated.
\end{enumerate}

To generate larger Yukawa couplings for the third generation, it was proposed in \cite{Babu:2006rp}
that one
include additional $10$ and $45$ of matter. These allow couplings at the cubic level of the
form
\beqn\label{1.6}
h^{(10)}_{\acute{a}\acute{b}} 16_{\acute{a}}\cdot144_{\acute{b}}\cdot10, ~~~h^{(45)}_{\acute{a}\acute{b}} 16_{\acute{a}}\cdot\overline{144}_{\acute{b}}\cdot45,
\eeqn
where $\acute{a}, \acute{b}$ are the generation indices and where
 the $SU(5)\times U(1)$ decomposition of $45$ plet is given in Eq.(\ref{1.1}), while the
$SU(5)\times U(1)$ decomposition of $10$ plet is given by
\beqn \label{1.7}
10= \bar 5(-2)+ 5(2).
\eeqn
In this case one can
define the combination of $16_{\acute{a}}$ that couples with the 10 plet and the 45 plet of
matter as the third generation. The production  of the third generation fermion masses depends on the
choice of the light Higgs doublets. For the light Higgs  doublets with $U(1)$ quantum numbers
$\pm 3$ mass generation occurs for the b quark, the tau lepton and for the top quark. However,
no Dirac mass is generated for the neutrino and thus a neutrino mass generation from a Type I See Saw
cannot occur. For another choice of the  light Higgs doublets where the $U(1)$ quantum numbers for
the doublets are $\pm 7$, one finds that the b quark, and the tau lepton get masses and a Dirac mass is
generated for the neutrino, but there is no generation of mass for the top quark.
In short it is not possible with a $10$ plet and a $45$ plet of matter to give masses to the
up and  to the down quarks,  to the charged lepton and also give a Dirac mass to the neutrino all at once.
 To overcome this problem we  analyze in this
paper the alternate possibility of including a $45$ plet and a $120$ plet of matter fields.  Thus as an alternative
to Eq.(\ref{1.6}) we will consider the cubic couplings
\beqn\label{1.8}
h^{(120)}_{\acute{a}\acute{b}} 16_{\acute{a}}\cdot144_{\acute{b}}\cdot120, ~~~h^{(45)}_{\acute{a}\acute{b}} 16_{\acute{a}}\cdot\overline{144}_{\acute{b}}\cdot45,
\eeqn
where the $SU(5)\times U(1)$ decomposition of $120$ is given by
\beqn\label{1.9}
120= 5(2) +\bar 5(-2) +10(-6)+ \overline{10}(6)+45(2).
\eeqn

The outline of the rest of the paper is as follows: In Sec.(\ref{2}) we
discuss the Higgs sector of $SO(10)$ and the conditions for the spontaneous breaking of $SO(10)$,
 as well as the generation of light Higgs doublets. In Sec.(\ref{3})  we compute the
Yukawa couplings that arise from the
quartic interactions
between matter and Higgs fields.  Here  we also compute quark, charged lepton and neutrino masses
arising from the quartic couplings.
In Sec.(\ref{4}) we analyze the
 cubic couplings that arise from inclusion of $45_M$ and $120_M$ of matter.
In this section we then discuss the production of masses  for the third generation fermions from these couplings.
In Sec.(\ref{5}) we give an analysis of $b-\tau-t$  unification. In Sec.(\ref{6}) we  give an
analysis of the $\tau$ neutrino mass.  Proton stability is briefly  discussed in Sec.(\ref{7}). Conclusions are given in Sec.(\ref{8}). Further details of the analysis are given in several  Appendices which
are referred to in the main text of the paper. Thus
Appendix A gives  a brief discussion of the formalism and of the calculational techniques. Details of the
analysis of quartic couplings are given in Appendix B, and the general mass formulae arising from
the quartic couplings after spontaneous breaking are given in Appendix  C. Appendix D gives details of the
analysis of the cubic interactions $16\cdot45\cdot\overline{144}$ and $16\cdot120\cdot144$ while the details of the
couplings that generate fermionic masses from cubic couplings are given in Appendix E.  Appendix F gives
the definitions of various quantities that enter in the diagonalization of the $b.\tau, t$ mass matrices.
Finally Appendix G defines the Yukawa coupling parameters that enter in Sec.(4).

\section{The Higgs Sector and the Spontaneous Breaking of $SO(10)$\label{2}}
As discussed in \cite{Babu:2005gx,Babu:2006rp}
spontaneous symmetry breaking  in the Higgs sector
can be  generated by considering the superpotential
\beqn
 {\mathsf W}&=& M {(\overline{144}_H\cdot 144_H)} + \frac{1}{M'}
 \left[ \lambda_{45_1}{ (\overline{144}_H\cdot
144_H)_{45_1}
 (\overline{144}_H\cdot 144_H)_{45_1}}\right.
 \nonumber\\
 &&\left.+  \lambda_{45_2}  { (\overline{144}_H\cdot
144_H)_{45_2}
 (\overline{144}_H\cdot 144_H)_{45_2}}
 +  \lambda_{210} { (\overline{144}_H\cdot
144_H)_{210}
 (\overline{144}_H\cdot 144_H)_{210}}\right].
 \label{2.1}
  \eeqn
We look for solutions where the $24$ plet of $SU(5)$  ${\bf Q}^i_{j}$
 in the $144$
and the 24-plet of $SU(5)$   ${\bf P}^i_{j}$ in
$\overline{144}$ (see Appendix A) develop
VEVs which preserve the Standard Model gauge group $SU(3)_C\times SU(2)_L\times U(1)Y$ such that
\begin{eqnarray}
 <{\bf Q}^i_{j}>= q~\textnormal{diag}(2,2,2,-3,-3),~~ <{\bf P}^i_{j}>= p~\textnormal{diag}(2,2,2,-3,-3),
\label{2.2}
\end{eqnarray}
which results in the minimization condition for the potential so that
\begin{eqnarray}\label{symmetrybreaking}
\frac{MM'}{qp}=116\lambda_{45_{_1}}+7\lambda_{45_{_2}}
+4\lambda_{210}.
\label{2.3}
\end{eqnarray}
A fine tuning is needed to get a pair of light Higgs doublets so one is working within the framework
of a landscape type scenario. As discussed in \cite{Babu:2006rp} a light pair of
 Higgs doublets can be gotten by looking at  the mass
matrices for the Higgs doublets ${\bf Q}_a,  {\bf\widetilde Q}_a, {\bf P}_a, {\bf P}^a, {\bf\widetilde P}^a, {\bf Q}^a$
(see Appendix A).
 Here the $SU(2)$ index $a$ can take on values 4 and 5.

We will denote the fields in the mass diagonal basis with a prime and these are related to the
unprimed fields by a unitary rotation given by

\begin{eqnarray}\label{rotatedhiggs}
 \left[\matrix{({\bf { Q}}^{\prime
}_{a},{\bf { P}}^{\prime a} )\cr ({\bf {\widetilde
Q}}^{\prime}_{a},{\bf {\widetilde P}}^{\prime a}) }\right]& =&
\left[\matrix{ \cos\vartheta_{\mathsf D} & \sin\vartheta_{\mathsf
D}\cr -\sin\vartheta_{\mathsf D} & \cos\vartheta_{\mathsf
D}}\right]\left[\matrix{({\bf{ Q}}_{a}, {\bf { P}}^{a})\cr ({\bf
{\widetilde Q}}_{ a},{\bf {\widetilde P}}^{ a})
 }\right],
 \label{2.4}
\end{eqnarray}
where the rotation angle $\vartheta_D$ is given by
\begin{eqnarray}
\tan\vartheta_{\mathsf D}&=&\frac{1}{{\mathsf d_3}}\left({\mathsf
d_2}+\sqrt{{\mathsf d_2}^2+{\mathsf d_3}^2}\right),
\label{2.5}
\end{eqnarray}
and where
\begin{eqnarray}
{\mathsf
d}_1&=&-\frac{2}{5}M+\frac{qp}{M'}\left(\frac{296}{5}\lambda_{45_1}
-16\lambda_{45_2}-\frac{392}{15}\lambda_{210}\right),\nonumber\\
{\mathsf
d}_2&=&-\frac{8}{5}M+\frac{qp}{M'}\left(-\frac{1036}{5}\lambda_{45_1}
+\frac{1}{2}\lambda_{45_2}+\frac{427}{15}\lambda_{210}\right),\nonumber\\
{\mathsf
d_3}&=&2\sqrt{\frac{3}{5}}\frac{qp}{M'}\left(10\lambda_{45_1}
+\frac{5}{4}\lambda_{45_2}-\frac{5}{6}\lambda_{210}\right).
\label{2.6}
\end{eqnarray}
The diagonalization of the mass matrix gives three Higgs doublets with
 mass eigenvalues given by
\begin{eqnarray}\label{doublets}
M_{{\mathsf
D}_1}&=&M+\frac{qp}{M'}(180\lambda_{45_1}+9\lambda_{45_2}-10\lambda_{210}),\nonumber\\
M_{{\mathsf D}_2,{\mathsf D}_3}&=&\frac{1}{2}\left({\mathsf
d_1}\pm\sqrt{{\mathsf d_2}^2+{\mathsf d_3}^2}\right).
\label{2.7}
\end{eqnarray}
With a constraint one can choose any one  of the three doublets to be light and
others super heavy. The light doublets that emerge are defined precisely in Eq.(\ref{3.7})
and are  labeled ${\mathsf D_1}$, ${\mathsf D_2}$, and ${\mathsf D_3}$.

\section{Yukawa  Couplings for Fermions from Quartic Interactions \label{3}}
As mentioned already Yukawa couplings can arise from the quartic interactions with two
matter and two Higgs fields of the type $16\cdot16\cdot144\cdot144$.  Specifically we consider below
the following possibilities, namely
\beqn\label{3.0}
\begin{array} {ccc}
~\left\{\zeta_{\acute{a}\acute{b},\acute{c}\acute{d}}^{^{(10)(+)}}\right\}~\left(16_{\acute{a}}\cdot
16_{\acute{b}}\right)_{10}\left(144_{\acute{c}}\cdot
144_{\acute{d}}\right)_{10}&&
~\left\{\xi_{\acute{a}\acute{b},\acute{c}\acute{d}}^{^{(10)(+)}}\right\}~\left(16_{\acute{a}}\cdot
16_{\acute{b}}\right)_{10}\left(\overline{144}_{\acute{c}}
\cdot \overline{144}_{\acute{d}}\right)_{10}\nonumber\\
~\left\{\varrho_{\acute{a}\acute{b},\acute{c}\acute{d}}^{^{(126,\overline{126})(+)}}\right\}~\left(16_{\acute{a}}\cdot
16_{\acute{b}}\right)_{\overline{126}} \left(144_{\acute{c}}\cdot
144_{\acute{d}}\right)_{126}&&
~\left\{\lambda_{\acute{a}\acute{b},\acute{c}\acute{d}}^{^{(45)}}\right\}~\left(16_{\acute{a}}\cdot
\overline{144}_{\acute{b}}\right)_{45} \left(16_{\acute{c}}\cdot \overline{144}_{\acute{d}}\right)_{45},\nonumber\\
~\left\{\zeta_{\acute{a}\acute{b},\acute{c}\acute{d}}^{^{(120)(-)}}\right\}~\left(16_{\acute{a}}\cdot
16_{\acute{b}}\right)_{120}\left({144}_{\acute{c}}\cdot{144}_{\acute{d}}\right)_{120}&&
~\left\{\xi_{\acute{a}\acute{b},\acute{c}\acute{d}}^{^{(120)(-)}}\right\}~\left(16_{\acute{a}}\cdot
16_{\acute{b}}\right)_{{120}}
\left(\overline{144}_{\acute{c}}\cdot \overline{144}_{\acute{d}}\right)_{120}\nonumber\\
~\left\{\lambda_{\acute{a}\acute{b},\acute{c}\acute{d}}^{^{(54)}}\right\}~\left(16_{\acute{a}}\cdot
\overline{144}_{\acute{b}}\right)_{54} \left(16_{\acute{c}}\cdot
\overline{144}_{\acute{d}}\right)_{54}&&
~\left\{\lambda_{\acute{a}\acute{b},\acute{c}\acute{d}}^{^{(10)}}\right\}~\left(16_{\acute{a}}\cdot
144_{\acute{b}}\right)_{10}\left(16_{\acute{c}}\cdot
144_{\acute{d}}\right)_{10}
\label{3.0}
\end{array}
\eeqn
Here the quantities
within the \{\} are the coupling  constants associated with the particular
quartic coupling\cite{Nath:2005bx}. After spontaneous breaking of $SO(10)$  at the GUT scale these
interactions generate Yukawa couplings for the quarks and leptons. Also they lead to interactions
appropriate for the generation of Dirac type neutrino masses as well as Majorana masses allowing for
a See Saw mechanism\cite{seesaw} to operate.  A detailed analysis of these quartic couplings is given in Appendix B.
 We give here some relevant results.\\

For the down-type quarks and for the charged leptons, the quartic interactions listed in the beginning of this section
lead to the
following set of Yukawa  couplings which arise when the $24$ plets in $144$ and $\overline{144}$ acquire
VEVs:
\begin{eqnarray}\label{3.1}
&{\bf\widehat M}_{\acute{a}}^{ij}{\bf\widehat M}_{\acute{b}j}&\left\{2\left[
-\lambda_{\acute{a}\acute{c},\acute{b}\acute{d}}^{^{(45)}}
-\lambda_{\acute{a}\acute{c},\acute{b}\acute{d}}^{^{(54)}}
+8\xi_{\acute{a}\acute{b},\acute{c}\acute{d}}^{^{(10)(+)}}+\frac{8}{3}\xi_{\acute{a}\acute{b},\acute{c}\acute{d}}^{^{(120)(-)}}\right]{{\bf\widehat P}}_{\acute{c}i}^k{{\bf\widehat P}}_{\acute{d}k}\right.\nonumber\\
&&\left.+\frac{1}{\sqrt
5}\left[\frac{1}{15}\varrho_{\acute{a}\acute{b},\acute{c}\acute{d}}^{^{(126,\overline{126})(+)}}
+8\zeta_{\acute{a}\acute{b},\acute{c}\acute{d}}^{^{(10)(+)}}
+\frac{8}{3}\zeta_{\acute{a}\acute{b},\acute{c}\acute{d}}^{^{(120)(-)}}\right]{{\bf\widehat Q}}_{\acute{c}k}{{\bf\widehat Q}}_{\acute{d}j}^k \right.\nonumber\\
&&\left.+2\left[\frac{1}{15}\varrho_{\acute{a}\acute{b},\acute{c}\acute{d}}^{^{(126,\overline{126})(+)}}
+8\zeta_{\acute{a}\acute{b},\acute{c}\acute{d}}^{^{(10)(+)}}
-\frac{8}{3}\zeta_{\acute{a}\acute{b},\acute{c}\acute{d}}^{^{(120)(-)}}\right]{{\bf\widehat Q}}_{\acute{c}ik}^l{{\bf\widehat Q}}_{\acute{d}l}^k\right\}\nonumber\\
&+{\bf\widehat M}_{\acute{a}}^{ij}{\bf\widehat M}_{\acute{b}k}&\left\{\frac{1}{\sqrt
5}\left[\frac{4}{15}\varrho_{\acute{a}\acute{b},\acute{c}\acute{d}}^{^{(126,\overline{126})(+)}}
-\lambda_{\acute{a}\acute{c},\acute{b}\acute{d}}^{^{(10)}}
+\frac{16}{3}\zeta_{\acute{a}\acute{b},\acute{c}\acute{d}}^{^{(120)(-)}}\right]{\bf\widehat Q}_{\acute{c}j}{{\bf\widehat Q}}_{\acute{d}i}^k \right.\nonumber\\
&&\left.+\frac{4}{
3}\left[-\frac{1}{5}\varrho_{\acute{a}\acute{b},\acute{c}\acute{d}}^{^{(126,\overline{126})(+)}}
+4\zeta_{\acute{a}\acute{b},\acute{c}\acute{d}}^{^{(120)(-)}}\right]{\bf\widehat Q}_{\acute{c}ij}^l{{\bf\widehat Q}}_{\acute{d}l}^k \right.
\left.+2\left[-\lambda_{\acute{a}\acute{c},\acute{b}\acute{d}}^{^{(45)}}
+\lambda_{\acute{a}\acute{c},\acute{b}\acute{d}}^{^{(54)}}\right]
{\bf\widehat P}_{\acute{c}j}^k{{\bf\widehat P}}_{\acute{d}i}\right\},
\end{eqnarray}
where
${\bf\widehat M}_{\acute{a}}^{ij}$ is the $10$ plet of $SU(5)$  matter fields and
${\bf\widehat M}_{\acute{b}j}$ is the $\bar 5$ of $SU(5)$ matter fields and $\acute{a},\acute{b}$
are the generation indices (see Appendix A).
For the up-type quarks
the quartic interactions listed in the beginning of this section
 lead to the
following set of cubic  couplings which again arise when the $24$ plets in $144$ and $\overline{144}$ develop
VEVs:
\begin{eqnarray}\label{3.2}
&\epsilon_{ijklm}{\bf\widehat M}_{\acute{a}}^{ij}{\bf\widehat M}_{\acute{b}}^{kl}&\left\{2\left[
\frac{2}{15}\varrho_{\acute{a}\acute{b},\acute{c}\acute{d}}^{^{(126,\overline{126})(+)}}
+\zeta_{\acute{a}\acute{b},\acute{c}\acute{d}}^{^{(10)(+)}}+\frac{2}{3}\zeta_{\acute{a}\acute{b},\acute{c}\acute{d}}^{^{(120)(-)}}\right]{{\bf\widehat Q}}_{\acute{c}}^n{{\bf\widehat Q}}_{\acute{d}n}^m\right.\nonumber\\
&&\left.+2\left[\xi_{\acute{a}\acute{b},\acute{c}\acute{d}}^{^{(10)(+)}}\right]{\bf\widehat P}_{\acute{c}n}^p{{\bf\widehat P}}_{\acute{d}p}^{nm} -\frac{1}{\sqrt
5}\left[\xi_{\acute{a}\acute{b},\acute{c}\acute{d}}^{^{(10)(+)}}\right]{\bf\widehat P}_{\acute{c}n}^m{{\bf\widehat P}}_{\acute{d}}^{n}\right\}\nonumber\\
&+\epsilon_{ijklm}{\bf\widehat M}_{\acute{a}}^{in}{\bf\widehat M}_{\acute{b}}^{jk}&\left\{\left[-\frac{1}{2}\lambda_{\acute{a}\acute{c},\acute{b}\acute{d}}^{^{(45)}}
-\frac{1}{2}\lambda_{\acute{a}\acute{c},\acute{b}\acute{d}}^{^{(54)}}
+\frac{4}{3}\xi_{\acute{a}\acute{b},\acute{c}\acute{d}}^{^{(120)(-)}}\right]{\bf\widehat P}_{\acute{c}n}^p{{\bf\widehat P}}_{\acute{d}p}^{lm}+\frac{1}{\sqrt
5}\left[-\lambda_{\acute{a}\acute{c},\acute{b}\acute{d}}^{^{(45)}}+\frac{4}{3}
\xi_{\acute{a}\acute{b},\acute{c}\acute{d}}^{^{(120)(-)}} \right]
{\bf\widehat P}_{\acute{c}n}^l{{\bf\widehat P}}_{\acute{d}}^{m}\right\}\nonumber\\
&+\frac{1}{2}\epsilon_{ijklm}{\bf\widehat M}_{\acute{a}}^{np}{\bf\widehat M}_{\acute{b}}^{ij}&\left[\lambda_{\acute{a}\acute{c},\acute{b}\acute{d}}^{^{(45)}}
-\lambda_{\acute{a}\acute{c},\acute{b}\acute{d}}^{^{(54)}}\right]{\bf\widehat P}_{\acute{c}p}^k{{\bf\widehat P}}_{\acute{d}n}^{lm}.
\end{eqnarray}
 The quartic interactions listed in the beginning of this section lead to the following interactions which
 generate a Dirac mass for the neutrino
\begin{eqnarray}\label{3.3}
&{\bf\widehat M}_{\acute{a}i}{\bf\widehat M}_{\acute{b}}&\left\{\frac{1}{\sqrt
5}\left[
-3\lambda_{\acute{a}\acute{c},\acute{b}\acute{d}}^{^{(45)}}
+\lambda_{\acute{a}\acute{c},\acute{b}\acute{d}}^{^{(54)}}
+8\xi_{\acute{a}\acute{b},\acute{c}\acute{d}}^{^{(10)(+)}}+\frac{8}{3}\xi_{\acute{a}\acute{b},\acute{c}\acute{d}}^{^{(120)(-)}}\right]{{\bf\widehat P}}_{\acute{c}}^j{{\bf\widehat P}}_{\acute{d}j}^i\right.\nonumber\\
&&\left.+2\left[
-\lambda_{\acute{a}\acute{c},\acute{b}\acute{d}}^{^{(45)}}
-\lambda_{\acute{a}\acute{c},\acute{b}\acute{d}}^{^{(54)}}
+8\xi_{\acute{a}\acute{b},\acute{c}\acute{d}}^{^{(10)(+)}}+\frac{8}{3}\xi_{\acute{a}\acute{b},\acute{c}\acute{d}}^{^{(120)(-)}}\right]{{\bf\widehat P}}_{\acute{c}k}^{ij}{{\bf\widehat P}}_{\acute{d}j}^k\right.\nonumber\\
&&\left.+2\left[\frac{16}{5}\varrho_{\acute{a}\acute{b},\acute{c}\acute{d}}^{^{(126,\overline{126})(+)}}
+\lambda_{\acute{a}\acute{d},\acute{b}\acute{c}}^{^{(10)}}
-8\zeta_{\acute{a}\acute{b},\acute{c}\acute{d}}^{^{(10)(+)}}
+\frac{8}{3}\zeta_{\acute{a}\acute{b},\acute{c}\acute{d}}^{^{(120)(-)}}\right]{{\bf\widehat Q}}_{\acute{c}}^{j}{{\bf\widehat Q}}_{\acute{d}j}^i\right \}.
\end{eqnarray}\\
Here ${\bf\widehat M}_{\acute{b}}$ is the SM singlet field (see Appendix A).
 Dirac masses arise from the above interactions when $24$ plets of fields
${{\bf\widehat Q}}_{j}^i$  and  ${{\bf\widehat P}}_{j}^i$ develop VEVs followed by the fields
${{\bf\widehat P}}^j$ and ${{\bf\widehat Q}}^j$ etc. developing VEVs.
Next we exhibit the quartic couplings that  produce the RR Majorana masses that enter in Type I See Saw.
These are
\begin{eqnarray}\label{3.4}
&{\bf\widehat M}_{\acute{a}}{\bf\widehat M}_{\acute{b}}&\left\{\left[
-\lambda_{\acute{a}\acute{c},\acute{b}\acute{d}}^{^{(45)}}
+\lambda_{\acute{a}\acute{c},\acute{b}\acute{d}}^{^{(54)}}
\right]{{\bf\widehat P}}_{\acute{c}j}^i{{\bf\widehat P}}_{\acute{d}i}^j
+\frac{4}{\sqrt 5}\left[
\frac{4}{15}\varrho_{\acute{a}\acute{b},\acute{c}\acute{d}}^{^{(126,\overline{126})(+)}}+
\lambda_{\acute{a}\acute{c},\acute{b}\acute{d}}^{^{(10)}}
\right]{{\bf\widehat Q}}_{\acute{c}}^i{{\bf\widehat Q}}_{\acute{d}i}\right\}.
\end{eqnarray}\\
  Since the first set of terms involve  two factors of $24$,  the  RR Majorana masses  are
  automatically of the GUT mass size.
Next we exhibit the quartic couplings that lead to  Type II See Saw masses. These are
\begin{eqnarray}\label{3.5}
&{\bf\widehat M}_{\acute{a}i}{\bf\widehat M}_{\acute{b}j}&\left[
\frac{1}{4}\left(\lambda_{\acute{a}\acute{c},\acute{b}\acute{d}}^{^{(45)}}
-\frac{201}{25}\lambda_{\acute{a}\acute{c},\acute{b}\acute{d}}^{^{(54)}}
\right)-\frac{1}{4}\left(5\lambda_{\acute{a}\acute{d},\acute{b}\acute{c}}^{^{(45)}}
-\frac{9}{5}\lambda_{\acute{a}\acute{d},\acute{b}\acute{c}}^{^{(54)}}
\right)-\frac{32}{15}\xi_{\acute{a}\acute{b},\acute{c}\acute{d}}^{^{(120)(-)}}
\right]{{\bf\widehat P}}_{\acute{c}}^{i}{{\bf\widehat P}}_{\acute{d}}^j.
\end{eqnarray}
When ${{\bf\widehat P}}^j$ develop VEVs of electroweak size they will lead to Type II See Saw masses
of the right size, i.e., of size $O(M_{EW}^2/\Lambda)$.

The analysis above was general including also the possibility of
 several generations for the
Higgs multiplets.  However, in the following we will assume that  there is  only one generation
for the $144+\overline{144}$ of Higgs.  Now the couplings
$\xi_{\acute{a}\acute{b},\acute{c}\acute{d}}^{^{(120)(-)}}$ and
 $\zeta_{\acute{a}\acute{b},\acute{c}\acute{d}}^{^{(120)(-)}}$  are anti-symmetric in the
 generation indices for the matter fields, i.e., the first two indices, as well as anti-symmetric in the
 last two indices for the Higgs fields. Since we limit ourselves to one generation of the  Higgs fields
 these couplings vanish, i.e.,
 \beqn\label{3.6}
\xi_{\acute{a}\acute{b},\acute{c}\acute{d}}^{^{(120)(-)}} =0=
 \zeta_{\acute{a}\acute{b},\acute{c}\acute{d}}^{^{(120)(-)}}.
 \eeqn
 With the quartic interactions  the  mass growth for the quark, lepton and
neutrino masses can occur in a variety of ways.
Thus  in the one generation of $144+\overline{144}$ of Higgs case the following distinct doublets occur
\begin{eqnarray}\label{3.7}
{\mathsf D}_1:({\bf Q}^{a}, {\bf P}_{a}),~~~~{\mathsf D}_2:({\bf
Q}_{a}^{\prime},{\bf P}^{\prime a}),~~~~{\mathsf
D}_3:({\bf{\widetilde Q}}_{a}^{\prime}, {\bf {\widetilde
P}}^{\prime a }),
\end{eqnarray}
where  $({\bf Q}_{a}^{\prime},{\bf P}^{\prime a})$ and
$({\bf{\widetilde Q}}_{a}^{\prime}, {\bf {\widetilde P}}^{\prime a
})$ are the rotated fields obtained after diagonalization of the
Higgs doublet mass matrix.
We begin by identifying $SU(3)_C\times
U(1)_{em}$ conserving VEV's
\begin{eqnarray}\label{3.8}
{<{\bf Q}_{\acute{c}j}^{i}>\choose {<\bf
P}_{\acute{c}j}^{i}>}&=&{q_{\acute{c }}\choose p_{\acute{c
}}}diag(2,2,2,-3,-3),\nonumber\\
<{\bf { Q}}_{\acute{d}5}>&=&<{{\bf  Q}}_{{\acute
{d}}5}^{\prime}>\cos\vartheta_{\mathsf D}+<{\widetilde{\bf
Q}}_{{\acute {d}}5}^{\prime}>\sin\vartheta_{\mathsf D},\nonumber\\
 <{\bf  {
P}}_{\acute{d}}^{5}
>&=&<{{\bf  P}}_{{\acute
{d}}}^{\prime 5}>\cos\vartheta_{\mathsf D}+<{\widetilde{\bf
P}}_{{\acute
{d}}}^{\prime 5}>\sin\vartheta_{\mathsf D},\nonumber\\
<{\bf{\widetilde Q}}_{\acute {d}5}>&=&-<{{\bf Q}}_{{\acute
{d}}5}^{\prime}>\sin\vartheta_{\mathsf D}+<{\widetilde{\bf
Q}}_{{\acute
{d}}5}^{\prime}>\cos\vartheta_{\mathsf D},\nonumber\\
     <{\bf{\widetilde P}}_{\acute {d}}^5>&=&-<{{\bf
P}}_{{\acute {d}}}^{\prime 5}>\sin\vartheta_{\mathsf
D}+<{\widetilde{\bf P}}_{{\acute {d}}}^{\prime
5}>\cos\vartheta_{\mathsf D}.
\end{eqnarray}
We now compute the down quark ($M^{d}$), charged lepton
($M^{e}$), up quark ($M^{u}$), Dirac neutrino
($M^{\nu D}$), RR type neutrino ($M^{RR}$) and LL type
neutrino ($M^{LL}$) mass matrices in terms of the primitive mass parameters. The following mass terms appear in the
Lagrangian:
\begin{eqnarray}\label{3.9}
{\mathsf L }^{mass}&=&M^{u}_{\acute {a}\acute
{b}}~~{\overline {\bf{U}}}_{R\acute {a}\alpha}~{\bf U}^{\alpha
}_{L\acute {b}}+
M^{d}_{\acute {a}\acute
{b}}~~{\overline
{\bf{D}}}_{R\acute {a}\alpha}~{\bf D}^{\alpha}_{L\acute {b}}+
M^{e}_{\acute {a}\acute {b}}~~
{\overline {\bf{E}}}_{R\acute {a}}^{(-)\mathsf c}~{\bf E}^{(+)
}_{L\acute {b}}\nonumber\\
&&+M^{\nu D}_{\acute {a}\acute
{b}}~~{\overline {\Nu}}_{R\acute {a}}~{\Nu}_{L\acute {b}}
+M^{RR}_{\acute {a}\acute
{b}}~~{\overline {\Nu}}_{R\acute {a}}~{\Nu}_{L\acute {b}}^{\mathsf c}
+M^{LL}_{\acute {a}\acute {b}}~~{\overline
{\Nu}}_{R\acute {a}}^{\mathsf c}~{\Nu}_{L\acute {b}}+H.c..
\end{eqnarray}
The general expressions for the mass matrices $M^{u}, M^{d}$ etc. are given
in Appendix C.
We consider one at a time each of the doublets ${\mathsf
D}_1$, ${\mathsf D}_2$ and ${\mathsf D}_3$  being massless.
\\\noindent\textsc{{\bf Case I:} ~ massless doublet~~${\mathsf
D}_1$}: Here
the down quark and the charged lepton masses are  given by
\beqn\label{3.10}
M^d=X+Y, ~~M^e=X-3Y,
\label{xy}
\eeqn
where
\begin{eqnarray}
X=\frac{1}{2}B^{(10)}_{5}-\frac{3}{4}B^{(45)}+\frac{3}{4}B^{(54)},
~~Y=\frac{1}{2}B^{(10)}_{5}+\frac{7}{4}B^{(45)}+\frac{1}{4}B^{(54)},\label{3.11}\\
\nonumber\\
M^{u}=A^{(10)}_{1}+A^{(126)},
~~M^{D\nu}=A^{(10)}_{1}+A^{(10)}_6-3A^{(126)},\label{3.12}\\
\nonumber\\
M^{RR}=C^{(45)}+C^{(54)},~~M^{LL}=0.\label{3.13}
 \end{eqnarray}
An explicit computation of the terms $A,B,C$ is given in Table (1).

\noindent\textsc{{\bf Case II:} ~massless doublet~~${\mathsf
D}_2$}: For this case the expressions for $X,Y$ and for the other matrices
take the following form
\begin{eqnarray}
 X&=&\frac{1}{2}B^{(10)}_{1}+\frac{1}{2}B^{(10)}_{3}
+\frac{9}{8}B^{(10)}_{6} +\frac{21}{22}B^{(126)}_{1}
+\frac{3}{2}B^{(126)}_{3},\nonumber\\
Y&=&\frac{1}{2}B^{(10)}_{1}+\frac{1}{2}B^{(10)}_{3}
-\frac{1}{8}B^{(10)}_{6} +\frac{1}{22}B^{(126)}_{1}
-\frac{1}{2}B^{(126)}_{3}\label{3.14},\\
\nonumber\\
M^{u}&=&A^{(10)}_{2}+A^{(10)}_4+A^{(45)}_{1}+A^{(45)}_{3}+A^{(54)}_{1},\label{3.15}
\\
\nonumber\\
M^{D\nu}&=&A^{(10)}_{2}+A^{(10)}_4+\frac{9}{8}\left[1-\frac{5}{4}\frac{\lambda_{{\acute
{a}}{\acute {d}},{\acute {b}}{\acute
{c}}}^{^{(45)}}}{\lambda_{{\acute {a}}{\acute {c}},{\acute
{b}}{\acute
{d}}}^{^{(45)}}}\right]^{-1}A^{(45)}_{1}\nonumber\\
&&+\frac{15}{4}\left[1-\frac{9}{2}\frac{\lambda_{{\acute
{a}}{\acute {d}},{\acute {b}}{\acute
{c}}}^{^{(45)}}}{\lambda_{{\acute {a}}{\acute {c}},{\acute
{b}}{\acute
{d}}}^{^{(45)}}}\right]^{-1}A^{(45)}_{3}
+ \frac{15}{4}A^{(54)}_{1}+A^{(54)}_{3},\label{3.16}
\\
\nonumber\\
M^{RR}&=&C^{(45)}+C^{(54)},
~~M^{LL}=D^{(45)}_1+D^{(54)}_{1}.\label{3.17}
\end{eqnarray}
\\
\noindent\textsc{{\bf Case III:} ~massless doublet~~${\mathsf
D}_3$}: For the massless doublet ${\mathsf D_3}$ case the mass matrices take the form
\begin{eqnarray}
 X
&=&\frac{1}{2}B^{(10)}_{2}+\frac{1}{2}B^{(10)}_{4}
+\frac{9}{8}B^{(10)}_{7} +\frac{21}{22}B^{(126)}_{2}
+\frac{3}{2}B^{(126)}_{4},\nonumber\\
Y&=&\frac{1}{2}B^{(10)}_{2}+\frac{1}{2}B^{(10)}_{4}
-\frac{1}{8}B^{(10)}_{7} +\frac{1}{22}B^{(126)}_{2}
-\frac{1}{2}B^{(126)}_{4},\label{3.18}\\
\nonumber\\
M^{u}&=&A^{(10)}_{3}+A^{(10)}_5+A^{(45)}_{2}+A^{(45)}_{4}+A^{(54)}_{2},\label{3.19}
\\ \nonumber\\
M^{D\nu}&=&A^{(10)}_{3}+A^{(10)}_5+\frac{9}{8}\left[1-\frac{5}{4}\frac{\lambda_{{\acute
{a}}{\acute {d}},{\acute {b}}{\acute
{c}}}^{^{(45)}}}{\lambda_{{\acute {a}}{\acute {c}},{\acute
{b}}{\acute
{d}}}^{^{(45)}}}\right]^{-1}A^{(45)}_{2}+\frac{15}{4}\left[1-\frac{9}{2}\frac{\lambda_{{\acute
{a}}{\acute {d}},{\acute {b}}{\acute
{c}}}^{^{(45)}}}{\lambda_{{\acute {a}}{\acute {c}},{\acute
{b}}{\acute
{d}}}^{^{(45)}}}\right]^{-1}A^{(45)}_{4}\nonumber\\
&&+\frac{15}{4}A^{(54)}_{2}+A^{(54)}_{4},\label{3.20}
\\ \nonumber\\
M^{RR}&=&C^{(45)}+C^{(54)},
 ~~M^{LL}=D^{(45)}_2+D^{(54)}_{2}.\label{3.21}
\end{eqnarray}

\section{Yukawa Couplings from
$\mathbf{16_{{{M}}}} \cdot \mathbf{{\overline{144}}_{{{H}}}}\cdot
\mathbf{45_{{{M}}}}$ and $\mathbf{16_{{{M}}}} \cdot
\mathbf{{{144}}_{{{H}}}}\cdot \mathbf{120_{{{M}}}}$\label{4}}
Since the third generation masses are much larger than the masses from the first two
generations, it would appear that such masses are likely to arise from cubic couplings.
Now the $144$ or  $\overline{144}$ of Higgs fields cannot directly couple to the $16$ plet
of matter  fields at the cubic level. In \cite{Babu:2006rp} an interesting idea was proposed in that such cubic
couplings can arise if there is an additional $10$ plet  and a $45$ plet of matter.
As discussed in Sec.(\ref{1}) by inclusion of cubic couplings of Eq.(\ref{1.6}) it was possible to
generate the third generation masses for the quarks and  for the leptons much larger than the
masses for the first two generation quarks and charged leptons. However, with the interactions
of Eq.(\ref{1.6}) it was not possible to produce a Dirac mass for the neutrino which is needed
for the generation of a See Saw mechanism and at the same time have a non-vanishing top
quark mass.
 We discuss this issue now more explicitly. Thus the
$16\cdot 10\cdot 144$ type couplings give the following terms that can contribute to the
quark, charged lepton and neutrino masses indicating again the
light doublet ${\mathsf D}_1$, ${\mathsf D}_2$ or  ${\mathsf D}_3$ for which the specific interaction listed below is valid
(see Eq.(\ref{3.7}) and Eq.(\ref{3.8}) for definitions of ${\mathsf D}_1$, ${\mathsf D}_2$ and ${\mathsf D}_3$)

\beqn\label{4.1}
b,\tau:&&10_{16}(-1)\cdot \bar 5_{10}(-2)\cdot <\bar 5_{144}(3)>~({\mathsf D}_2,{\mathsf D}_3),\nonumber\\
&&10_{16}(-1)\cdot \bar 5_{10}(-2)\cdot < \overline{45}_{144}(3)>~({\mathsf D}_2,{\mathsf D}_3),\nonumber\\
{LR} -\nu:&&1_{16}(-5)\cdot \bar 5_{10}(-2)\cdot  <5_{144}(7)>~({\mathsf D}_1).
\eeqn
while the
$16\cdot 45\cdot \overline{144}$ type couplings give the following terms that can contribute to the
quark, charged lepton and neutrino masses
\beqn\label{4.2}
b,\tau:&&\bar 5_{16}(3)\cdot \bar 5_{45}(4)\cdot <\bar 5_{\overline{144}}(-7)>~({\mathsf D}_1),\nonumber\\
t: &&10_{16}(-1)\cdot  10_{45}(4)\cdot  <5_{\overline{144}}(-3)> ~({\mathsf D}_2,{\mathsf D}_3),\nonumber\\
&&10_{16}(-1)\cdot  10_{45}(4)\cdot  <{45}_{\overline{144}}(-3)>~({\mathsf D}_2,{\mathsf D}_3),\nonumber\\
&&10_{16}(-1)\cdot  10_{45}(4)\cdot  <{45}_{\overline{144}}(-3)>~({\mathsf D}_2,{\mathsf D}_3),\nonumber\\
&&10_{16}(-1)\cdot  10_{45}(4)\cdot  <{5}_{\overline{144}}(-3)>~({\mathsf D}_2,{\mathsf D}_3).
\eeqn
Now suppose ($\bar 5_{144}(3), 5_{\overline{144}}(-3)$) and  ($\overline{45}_{144}(3),{45}_{\overline{144}}(-3)$) contain the
 light doublets (i.e.,  the doublets
 ${\mathsf D_2}$ or ${\mathsf D_3}$, see Eq.(\ref{3.7})) that develop a VEV at the electroweak scale while the remaining doublets including the ones arising from
$5_{144}(7)$ and $\bar 5_{\overline{144}}(-7)$ are heavy and are integrated out of the low energy
theory. In this case we will have mass generation for the $b$-quark and for the $t$-quark but no Dirac
mass generation for the third generation neutrino from these  couplings.
Alternately suppose the doublets in  ($5_{144}(7), \bar 5_{\overline{144}}(-7)$) (i.e., the
 ${\mathsf D_1}$ doublets,  see Eq.(\ref{3.7}))  are light while the remaining doublets (${\mathsf D_2}, {\mathsf D_3}$) are heavy
 and can be integrated out. In this case one finds that the bottom quark and the tau neutrino
 get Dirac masses but the top does not acquire a mass. Thus one concludes that the interactions
 of $16\cdot 10\cdot 144$ and $16\cdot 45\cdot \overline{144}$  produce masses for
either  the bottom, the tau and the top, or for the bottom, the tau and the tau neutrino but not for all.
Specifically it is not possible to generate simultaneously mass  for the top as well as a Dirac mass for the
tau neutrino necessary for realizing a Type I See Saw mass for the tau neutrino.

Next we consider the couplings $16\cdot 120\cdot 144$ where we have a $120$ plet of matter
instead of a $10$ plet of matter.  Mass  growths for the bottom, tau,  top and for the neutrino
are as follows
\beqn\label{4.3}
b,\tau:&&10_{16}(-1)\cdot \bar 5_{120}(-2)\cdot <\bar 5_{144}(3)>~({\mathsf D}_2,{\mathsf D}_3),\nonumber\\
&&10_{16}(-1)\cdot \bar 5_{120}(-2)\cdot < \overline{45}_{144}(3)> ~({\mathsf D}_2,{\mathsf D}_3),\nonumber\\
          &&\bar 5_{16}(3)\cdot 10_{120}(-6)\cdot <\bar 5_{144}(3)>~({\mathsf D}_2,{\mathsf D}_3),\nonumber\\
                           &&\bar 5_{16}(3)\cdot 10_{120}(-6)\cdot < \overline{45}_{144}(3)>~({\mathsf D}_2,{\mathsf D}_3),\nonumber\\
                          t:&&10_{16}(-1)\cdot 10_{120}(-6)\cdot <5_{144} (7)> ~({\mathsf D}_1),\nonumber\\
 LR-\nu: && 1_{16}(-5)\cdot\bar 5_{120}(-2)\cdot <5_{144}(7)>~({\mathsf D}_1).
\eeqn
As before let us again
 suppose  that ($\bar 5_{144}(3), 5_{\overline{144}}(-3)$) and  ($\overline{45}_{144}(3), {45}_{\overline{144}}(-3)$) contain the
 light doublets (i.e.,  the doublets
 ${\mathsf D_2}$ or ${\mathsf D_3}$) which develop  a VEV
 while the remaining doublets including those from
$5_{144}(7)$ and $\bar 5_{\overline{144}}(-7)$ are heavy.
In this case from Eqs.(\ref{4.2}) and (\ref{4.3}) we find that
 $b$ and $t$ quarks and $\tau$ lepton develop Dirac masses but  the $\tau$ neutrino does not get a mass.  However,
consider next the case where the light doublets arise from
($5_{144}(7), \bar 5_{\overline{144}}(-7)$) (which is the light
 ${\mathsf D_1}$ case) are light while the remaining doublets (${\mathsf D_2}, {\mathsf D_3}$) are heavy
 and can be integrated out.  This time from Eq.(\ref{4.2}) we find that the $b$ quark and the $\tau$ lepton
 get a mass while from Eq.(\ref{4.3}) we find that the top quark and the $\tau$ neutrino get  masses.
 Thus with the cubic couplings $16\cdot 45\cdot \overline{144}$ and $16\cdot 120\cdot 144$
 and with the doublet ${\mathsf D_1}$ chosen to  be the light doublet, all the third generation fermions get masses
 with the neutrino getting a Type I See Saw mass.
 We discuss now the details of the analysis with $16\cdot 45\cdot \overline{144}$ and $16\cdot 120\cdot 144$
 couplings.

 The superpotential with the $45$ plet of matter is given by
 \begin{eqnarray}\label{4.41}
 {\mathsf W}^{(45)} &=&{\mathsf W}_{{{ mass}}}^{(45)}+{\mathsf W}^{16  \cdot {45}\cdot\overline{144}},
 \end{eqnarray}
 where,
\begin{eqnarray}\label{4.42}
{\mathsf W}_{{{ mass}}}^{(45)}=m_F^{(45)}\widehat { {\bf
F}}_{\mu\nu}^{(45)}\widehat { {\bf F}}_{\mu\nu}^{(45)},~~~{\mathsf W}^{16 \cdot {45}\cdot\overline{144}}= \frac{1}{2!}h_{\acute{a}\acute{b}}^{(45)}<{\widehat\Psi}_{(+)\acute{a}}^{*}|B\Gamma_{[\mu}|{\widehat\Upsilon}_{(+)\nu]}
 \widehat { { {\bf F}}}_{\acute{b}\mu\nu}^{(45)}.
\end{eqnarray}
Similarly the superpotential with $120$ plet of matter is given by
\begin{eqnarray}\label{4.51}
 {\mathsf W}^{(120)} &=&{\mathsf W}_{{{ mass}}}^{(120)}+{\mathsf W}^{16  \cdot {120}\cdot{144}},
 \end{eqnarray}
\begin{eqnarray}\label{4.52}
 {\mathsf W}_{{{ mass}}}^{(120)}=m_F^{(120)}\widehat { {\bf F}}_{\mu\nu\rho}^{(120)}\widehat { {\bf
F}}_{\mu\nu\rho}^{(120)},~~~{\mathsf W}^{16  \cdot {120}\cdot{144}}=\frac{1}{3!}
h_{\acute{a}\acute{b}}^{(120)}<{\widehat\Psi}_{(+)\acute{a}}^{*}|B\Gamma_{[\mu}\Gamma_{\nu}|{\widehat\Upsilon}_{(-)\rho]} \widehat { { {\bf F}}}_{\acute{b}\mu\nu\rho}^{(120)}.
\end{eqnarray}
Defining
\begin{equation}\label{4.6}
f_{\acute{a}\acute{b}}^{(.)}\equiv
ih_{\acute{a}\acute{b}}^{(.)};~~~f^{(.)}~ \textnormal{real},
\end{equation}
 the couplings in the superpotential that
can contribute  to the top, the bottom and the tau  Yukawa  couplings are given by
\begin{eqnarray}\label{4.7}
{\mathsf
W}^{(45)}&=&\frac{1}{\sqrt{10}}f^{(45)}_{\acute{a}\acute{b}}\epsilon_{ijklm}
{\bf\widehat M}_{\acute{a}}^{ij} {\bf\widehat P}^k\widehat { {\bf
F}}_{\acute{b}}^{(45)lm} + 2\sqrt
2f^{(45)}_{\acute{a}\acute{b}} {\bf\widehat M}^{ij}_{\acute{a}} {\bf\widehat P}_{i}^k\widehat { {\bf F}}_{\acute{b}jk}^{(45)}\nonumber\\
&& + \frac{1}{\sqrt
2}f^{(45)}_{\acute{a}\acute{b}}\epsilon_{ijklm}\widehat {\bf
M}_{\acute{a}}^{ij} {\bf\widehat P}_{n}^{kl}\widehat { {\bf
F}}_{\acute{b}}^{(45)mn} + m_F^{(45)}\widehat { {\bf
F}}^{(45)ij}\widehat { {\bf F}}_{ij}^{(45)} \nonumber\\
&&-2\sqrt
2f^{(45)}_{\acute{a}\acute{b}}\widehat {\bf M}_{\acute{ai}}
{\bf\widehat P}_j\widehat { {\bf F}}_{\acute{b}}^{(45)ij} -m_F^{(45)}\widehat { {\bf F}}^{(45)}\widehat
{ {\bf F}}^{(45)},
\end{eqnarray}
and by
\begin{eqnarray}
{\mathsf
W}^{(120)}&=&-\frac{16}{\sqrt{15}}f^{(120)}_{\acute{a}\acute{b}}\widehat
{\bf M}_{\acute{a}} {\bf\widehat Q}_i\widehat { {\bf
F}}_{\acute{b}}^{(120)i} -16\sqrt{3}f^{(120)}_{\acute{a}\acute{b}}\widehat
{\bf M}_{\acute{a}} {\bf\widehat Q}^i\widehat { {\bf F}}_{\acute{b}i}^{(120)}\nonumber\\
&& -\frac{8}{\sqrt{3}}f^{(120)}_{\acute{a}\acute{b}}\widehat
{\bf M}_{\acute{a}i} {\bf\widehat Q}_j^i\widehat { {\bf
F}}_{\acute{b}}^{(120)j} -8\sqrt{\frac{3}{5}}f^{(120)}_{\acute{a}\acute{b}}\widehat
{\bf M}_{\acute{a}i} {\bf\widehat Q}_j\widehat { {\bf F}}_{\acute{b}}^{(120)ij}\nonumber\\
&& -8\sqrt{\frac{3}{5}}f^{(120)}_{\acute{a}\acute{b}}\widehat
{\bf M}_{\acute{a}}^{ij} {\bf\widehat Q}_i\widehat { {\bf
F}}_{\acute{b}j}^{(120)} + \frac{16}{\sqrt{3}}f^{(120)}_{\acute{a}\acute{b}}\widehat
{\bf M}_{\acute{a}}^{ij} {\bf\widehat Q}_{ij}^k\widehat { {\bf F}}_{\acute{b}k}^{(120)}\nonumber\\
&& -\frac{2}{\sqrt{3}}\epsilon_{ijklm}f^{(120)}_{\acute{a}\acute{b}}\widehat
{\bf M}_{\acute{a}}^{ij} {\bf\widehat Q}^k\widehat { {\bf
F}}_{\acute{b}}^{(120)lm} + \frac{16}{\sqrt{3}}f^{(120)}_{\acute{a}\acute{b}}\widehat
{\bf M}_{\acute{a}}^{ij} {\bf\widehat Q}_i^k\widehat { {\bf F}}_{\acute{b}jk}^{(120)}\nonumber\\
 && + m_F^{(120)}\widehat { {\bf
F}}_{ij}^{(120)}\widehat { {\bf F}}^{(120)ij} + 2m_F^{(120)}\widehat { {\bf
F}}_{i}^{(120)}\widehat { {\bf F}}^{(120)i}.
\end{eqnarray}
A further analysis of these couplings is given in Appendices D and E.
From these  we can extract the mass matrices for the bottom quark, for the tau lepton
and for the top quark. We discuss these below.

\subsection{Bottom Quark Mass}
Using the computations given in Eq.(\ref{e1}) of Appendix E we display the mass matrix
for the bottom quark. We have
\begin{eqnarray}\label{4.9}
\matrix{ {}^{(10_{16})}\!{\Large {\textnormal b}}_{{
L}}^{\alpha}&~~ {}^{(10_{45})}\!{\Large {\textnormal b}}_{{
L}}^{\alpha} &~~ {}^{({{5}}_{120})}\!{\Large {\textnormal b}}_{{
L}}^{\alpha}& {}^{({{10}}_{120})}\!{\Large {\textnormal b}}_{{
L}}^{\alpha}}~~~\nonumber\\
 {\mathtt M}_{ {b}}~=~ \matrix{
{}^{({\overline {5}}_{16})}\!\overline{{\Large {\textnormal
b}}}_{{ R}\alpha}\cr \cr {}^{({\overline
{5}}_{120})}\!\overline{{\Large {\textnormal b}}}_{{ R}\alpha}\cr
\cr {}^{({\overline{10}}_{120})}\!\overline{{\Large {\textnormal
b}}}_{{ R}\alpha}\cr\cr
{}^{({\overline{10}}_{45})}\!\overline{{\Large {\textnormal
b}}}_{{ R}\alpha}}\left(\matrix{
 0 &{m_{{{b}}}}^{(45)} & m_D^{(120)} &{m_{{ {b}}}}^{(120)\prime\prime} \cr\cr
{ m_{{ {b}}}}^{(120)\prime}& 0 & -2m_F^{(120)}& 0 \cr\cr
 -m_D^{(120)} & 0 & 0& -2m_F^{(120)}\cr\cr
 m_D^{(45)} & -2m_F^{(45)} & 0& 0}
 \right),
  \end{eqnarray}

where
\begin{eqnarray}\label{4.10}
&{m_{{ {b}}}}^{(120)\prime} =8f^{(120)}_{33}
\left[\sqrt{\frac{3}{5}}<{\bf Q}_5>+\frac{2}{3}<\widetilde{\bf
Q}_5>\right],&\nonumber\\
&{m_{{ {b}}}}^{(120)\prime\prime}
=-8\sqrt{\frac{3}{5}}f^{(120)}_{33}<{\bf Q}_5>,~~~~~
{m_{{ {b}}}}^{(45)} =-2\sqrt {2}f^{(45)}_{33}<{\bf P}_5>,&\nonumber\\
&m_D^{(120)}=\frac{16}{\sqrt 3}f^{(120)}_{33}
q,~~~~~m_D^{(45)}=-2\sqrt 2f^{(45)}_{33} p.&
\end{eqnarray}
Note that ${\mathtt M}_{{{b}}}$ is asymmetric, hence it is
diagonalized by a biunitary transformation consisting of
 two $4\times 4$ orthogonal matrices, $U_{{{
{b}}}}$ and $V_{{ b}}$ satisfying
$$
U_{{ b}}{\mathtt M}_{{{b}}} V_{{b}}^{\bf T}={\textnormal
{diag}}\left(\lambda_{{ b}_1}~,~\lambda_{{ b}_2}~,~\lambda_{{
b}_3}~,~\lambda_{{ b}_4} \right).
$$
The matrices $U_{{{b}}}$ and $V_{{ {b}}}$ are such that their
columns are eigenvectors of matrices ${\mathtt M}_{{ {b}}}{\mathtt
M}_{{ {b}}}^{\bf T}$ and ${\mathtt M}_{{ {b}}}^{\bf T}{\mathtt
M}_{{ {b}}}$ respectively:
$$
U_{{{b}}}\left[{\mathtt M}_{{{b}}}{\mathtt M}_{{ {b}}}^{\bf
T}\right]U_{{{b}}}^{\bf T}={ {diag}}\left(\lambda_{{
b}_1}^2~,~\lambda_{{b}_2}^2~,~\lambda_{{ b}_3}^2~,~\lambda_{{
b}_4}^2 \right) =V_{{ {b}}}\left[{\mathtt M}_{{ {b}}}^{\bf
T}{\mathtt M}_{{ {b}}}\right]V_{{ {b}}}^{\bf T}.
$$
Further, the rotated  fields can be expressed in terms of the
 original fields through
\begin{eqnarray}\label{4.11}
\left(\matrix{\overline{{\Large {\textnormal b}}}_{{1 R}\alpha}\cr
\overline{{\Large {\textnormal b}}}_{{2 R}\alpha}\cr
\overline{{\Large {\textnormal b}}}_{{
3R}\alpha}\cr\overline{{\Large {\textnormal b}}}_{{ 4R}\alpha}
}\right)=U_{{ {b}}}\left(\matrix{{}^{({\overline
{5}}_{16})}\!\overline{{\Large {\textnormal b}}}_{{ R}\alpha}\cr
{}^{({\overline {5}}_{120})}\!\overline{{\Large {\textnormal
b}}}_{{ R}\alpha}\cr
{}^{({\overline{10}}_{120})}\!\overline{{\Large {\textnormal
b}}}_{{ R}\alpha}\cr
{}^{({\overline{10}}_{45})}\!\overline{{\Large {\textnormal
b}}}_{{ R}\alpha} }\right),~~~~~\left(\matrix{{\Large {\textnormal
b}}_{{ 1L}}^{\prime\alpha}\cr {\Large {\textnormal b}}_{{
2L}}^{\prime\alpha}\cr {\Large {\textnormal b}}_{{
3L}}^{\prime\alpha}\cr {\Large {\textnormal b}}_{{
4L}}^{\prime\alpha} }\right)=V_{{
{b}}}\left(\matrix{{}^{(10_{16})}\!{\Large {\textnormal b}}_{{
L}}^{\alpha}\cr {}^{(10_{45})}\!{\Large {\textnormal b}}_{{
L}}^{\alpha}\cr
 {}^{(5_{120})}\!{\Large {\textnormal b}}_{{
L}}^{\alpha} \cr {}^{(10_{120})}\!{\Large {\textnormal b}}_{{
L}}^{\alpha} } \right).
\end{eqnarray}
\noindent The mass terms in the Lagrangian are then given by
\begin{equation}\label{4.12}
\lambda_{{ b}_1}~ \overline{{\Large {\textnormal b}}}_{{1
R}\alpha} {\Large {\textnormal b}}_{{1
L}}^{\prime\alpha}~+~\lambda_{{ b}_2}~ \overline{{\Large
{\textnormal b}}}_{{2 R}\alpha}{\Large {\textnormal b}}_{{
2L}}^{\prime\alpha} ~+~\lambda_{{ b}_3} ~\overline{{\Large
{\textnormal b}}}_{{3 R}\alpha} {\Large {\textnormal b}}_{{
3L}}^{\prime\alpha}~+~\lambda_{{ b}_4}~ \overline{{\Large
{\textnormal b}}}_{{ 4R}\alpha} {\Large {\textnormal b}}_{{
4L}}^{\prime\alpha}.
\end{equation}
The $\det\left({\mathtt M}_{{ {b}}}{\mathtt M}_{{ {b}}}^{\bf
T}-\lambda_{{ b}}^2{\bf 1}\right)$ gives a quartic equation in
$\lambda_{{ b}}^2$. In the limit when ${m_{{
{b}}}}^{(120)\prime},~{m_{{
{b}}}}^{(120)\prime\prime},~{m_{{{b}}}}^{(45)}$ are small, the light
eigenvalue squared $\lambda_{{ b}_1}^2$ can be calculated from
\beqn\label{4.13}
\lambda_{{ b}_1}^2\approx \frac{\det\left({\mathtt
M}_{{{b}}}{\mathtt M}_{{ {b}}}^{\bf T}\right)}{\Lambda_{{
b}_2}^2\Lambda_{{ b}_3}^2\Lambda_{{ b}_4}^2},
\label{det}
\eeqn
where $\Lambda_{{ b}_2}^2$, $\Lambda_{{ b}_3}^2$ and $\Lambda_{{
b}_4}^2$ are the exact square of the non-vanishing
 eigenvalues of the matrix
${\mathtt M}_{{{b}}}{\mathtt M}_{{ {b}}}^{\bf T}|_{{
{m_{{{b}}}}^{(.)}}=0}$. Eq.(\ref{4.13}) leads to the following values for the b-quark
mass for the cases   $\mathsf{D_1}, \mathsf{D_2}, \mathsf{D_3}$.
\begin{eqnarray}\label{4.14}
{m_{{ {b}}}}^2\equiv\lambda_{{ b}_1}^2\approx \left\{\matrix{
\epsilon_b\times
\left(f^{(45)}_{33}<{\bf P}_5>\right)^2;~~~~{\mathsf D}_1
\cr\cr
\delta_b\times
\left\{\matrix{\left(f^{(120)}_{33}<{\bf{
Q}}_{5}^{\prime}>\left[3\sqrt{3}\cos\vartheta_{\mathsf
D}+\sqrt{5}\sin\vartheta_{\mathsf D}\right]\right)^2;&{\mathsf
D}_2
\cr
\left(f^{(120)}_{33}<{\bf{\widetilde
Q}}_{5}^{\prime}>\left[-3\sqrt{3}\sin\vartheta_{\mathsf
D}+\sqrt{5}\cos\vartheta_{\mathsf D}\right]\right)^2;&{\mathsf
D}_3,
}\right. }\right.
\end{eqnarray}
where ${\mathsf D_1}, {\mathsf D_2}, {\mathsf D_3}$ are as defined in Eq.(\ref{3.7}) and
where $\epsilon_b$ and $\delta_b$ are as defined in Appendix G.
The transformation matrices $U_b$, $V_b$ that enter in Eq.(\ref{4.11}) take the form
\begin{eqnarray}
U_{{ {b}}}=\left(\matrix{\cos\theta_{u {b}} &
-\sin\theta_{u{b}}&0&0 \cr \sin\theta_{u{b}} &\cos\theta_{u{b}}&
0&0\cr 0& 0&\cos\phi_{u {b}}&- \sin\phi_{u {b}} \cr
0&0&\sin\phi_{u {b}}&\cos\phi_{u {b}}}\right),~~~V_{{
{b}}}=\left(\matrix{b_1&0&1&1\cr b_2&0&b_{4-}&b_{4+}\cr 0&1&0&0\cr
b_3&0&b_{5-}&b_{5+}}\right),\label{4.15}
\end{eqnarray}
where
\begin{eqnarray}\label{4.16}
\tan\theta_{u
{b}}&=&4\sqrt{\frac{2}{3}}\frac{1}{y_{120}},
~\tan 2\phi_{u
{b}}=\frac{2m_D^{(45)}m_D^{(120)}}{m_D^{(45)^2}\left(1+y_{45}^2\right)-m_D^{(120)^2}\left(1+\frac{3}{32}y_{120}^2\right)},
\end{eqnarray}
and $b_i$ are given in Appendix F.
\subsection{Tau Lepton Mass}
Using the computations given in Eq.(\ref{e2}) of Appendix E we display the mass matrix
for the tau lepton. We have
\begin{eqnarray}\label{4.17}
\matrix{ {}^{(10_{16})}\!{{\overline{ {\Tau}}}}_{{ R}}&~~
{}^{(10_{45})}\!{\overline{ {\Tau}}}_{{ R}} &~~
{}^{({{5}}_{120})}\!{\overline{ {\Tau}}}_{{ R}}&
{}^{({{10}}_{120})}\!{\overline{ {\Tau}}}_{{
R}}}~~~\nonumber\\
 {\mathtt M}_{ {\tau}}~=~ \matrix{
{}^{({\overline {5}}_{16})}\!{{{ {\Tau}}}}_{{ L}}\cr \cr
{}^{({\overline {5}}_{120})}\!{{{ {\Tau}}}}_{{ L}}\cr \cr
{}^{({\overline{10}}_{120})}\!{{{ {\Tau}}}}_{{ L}}\cr\cr
{}^{({\overline{10}}_{45})}\!{{{ {\Tau}}}}_{{ L}}}\left(\matrix{
 0 &{m_{{{\tau}}}}^{(45)} & m_E^{(120)} &{m_{{ {\tau}}}}^{(120)\prime\prime} \cr\cr
{ m_{{ {\tau}}}}^{(120)\prime}& 0 & -2m_F^{(120)}& 0 \cr\cr
 4m_E^{(120)} & 0 & 0& -2m_F^{(120)}\cr\cr
 m_E^{(45)} & -2m_F^{(45)} & 0& 0}
 \right),
  \end{eqnarray}
where
\begin{eqnarray}\label{4.18}
&{m_{{ {\tau}}}}^{(120)\prime} =8f^{(120)}_{33}
\left[-\sqrt{\frac{3}{5}}<{\bf Q}_5>+2<\widetilde{\bf
Q}_5>\right],&\nonumber\\
&{m_{{ {\tau}}}}^{(120)\prime\prime}
=8\sqrt{\frac{3}{5}}f^{(120)}_{33}<{\bf Q}_5>,~~~~~
{m_{{ {\tau}}}}^{(45)} =2\sqrt {2}f^{(45)}_{33}<{\bf P}_5>,&\nonumber\\
&m_E^{(120)}=-8\sqrt{3}f^{(120)}_{33} q,~~~~~m_E^{(45)}=-12\sqrt
2f^{(45)}_{33} p.&
\end{eqnarray}

Again in the limit ${m_{{ {\tau}}}}^{(120)\prime},~{m_{{
{\tau}}}}^{(120)\prime\prime},~{m_{{ {\tau}}}}^{(45)}$ are small, the light
eigenvalue for the $\tau$ mass take the form for the three cases ${\mathsf D_1}, {\mathsf D_2}, {\mathsf D_3}$ as below
\begin{eqnarray}\label{4.19}
{m_{{ {\tau}}}}^2\equiv\lambda_{{ \tau}_1}^2\approx
\left\{\matrix{
\epsilon_{\tau}\times
\left(f^{(45)}_{33}<{\bf
P}_5>\right)^2;~~~~{\mathsf D}_1
\cr\cr
\delta_{\tau}\times
\left\{\matrix{\left(f^{(120)}_{33}<{\bf{
Q}}_{5}^{\prime}>\left[3\sqrt{3}\cos\vartheta_{\mathsf
D}+2\sqrt{5}\sin\vartheta_{\mathsf D}\right]\right)^2;&{\mathsf
D}_2
\cr
\left(f^{(120)}_{33}<{\bf{\widetilde
Q}}_{5}^{\prime}>\left[-3\sqrt{3}\sin\vartheta_{\mathsf
D}+2\sqrt{5}\cos\vartheta_{\mathsf D}\right]\right)^2;&{\mathsf
D}_3,
}\right. }\right.
\end{eqnarray}
where $\epsilon_{\tau}$  and $\delta_{\tau}$ are defined in Appendix G.
The rotated fields can now be expressed in terms of the primitive
ones
\begin{eqnarray}\label{4.20}
\left(\matrix{\overline{{\Tau}}_{{1 R}}\cr \overline{{\Tau}}_{{2
R}}\cr \overline{{\Tau}}_{{ 3R}}\cr\overline{{\Tau}}_{{ 4R}}
}\right)=V_{{
{\tau}}}\left(\matrix{{}^{(10_{16})}\!\overline{{\Tau}}_{{ R}}\cr
{}^{(10_{45})}\!\overline{{\Tau}}_{{ R}}\cr
{}^{(5_{120})}\!\overline{{\Tau}}_{{ R}}\cr
{}^{(10_{120})}\!\overline{{\Tau}}_{{ R}}
}\right),~~~~~\left(\matrix{{\Tau}_{{ 1L}}^{\prime}\cr {\Tau}_{{
2L}}^{\prime}\cr {\Tau}_{{ 3L}}^{\prime}\cr {\Tau}_{{
4L}}^{\prime} }\right)=U_{{ {\tau}}}\left(\matrix{{}^{({\overline
{5}}_{16})}\!{\Tau}_{{ L}}\cr {}^{({\overline
{5}}_{120})}\!{\Tau}_{{ L}}\cr
 {}^{({\overline {10}}_{120})}\!{\Tau}_{{
L}} \cr {}^{({\overline {10}}_{45})}\!{\Tau}_{{ L}} } \right),
\end{eqnarray}
and the  mass terms in the Lagrangian are
\begin{equation}\label{4.21}
\lambda_{{ \tau}_1}~ \overline{{\Large {\Tau}}}_{{1 R}} {\Large
{\Tau}}_{{1 L}}^{\prime}~+~\lambda_{{ \tau}_2}~ \overline{{\Large
{\Tau}}}_{{2 R}}{\Large {\Tau}}_{{ 2L}}^{\prime} ~+~\lambda_{{
\tau}_3} ~\overline{{\Large {\Tau}}}_{{3 R}} {\Large {\Tau}}_{{
3L}}^{\prime}~+~\lambda_{{ \tau}_4}~ \overline{{\Large {\Tau}}}_{{
4R}} {\Large {\Tau}}_{{ 4L}}^{\prime}.
\end{equation}
Rotation matrices  $U_{\tau}$ and $V_{\tau}$ that enter in Eq.(\ref{4.20})
are given by
\begin{eqnarray}\label{4.22}
U_{{ {\tau}}}=\left(\matrix{\cos\theta_{u {\tau}} &
-\sin\theta_{u{\tau}}&0&0 \cr \sin\theta_{u{\tau}}
&\cos\theta_{u{\tau}}& 0&0\cr 0& 0&\cos\phi_{u {\tau}}&-
\sin\phi_{u {\tau}} \cr 0&0&\sin\phi_{u {\tau}}&\cos\phi_{u
{\tau}}}\right),~~~V_{{ {\tau}}}=\left(\matrix{\tau_1&0&1&1\cr
\tau_2&0&\tau_{4-}&\tau_{4+}\cr 0&1&0&0\cr
\tau_3&0&\tau_{5-}&\tau_{5+}}\right),
\end{eqnarray}
where
\begin{eqnarray}\label{4.23}
\tan\theta_{u{\tau}}=-2\sqrt{6}\frac{1}{y_{120}},
~\tan 2\phi_{u
{\tau}}=\frac{2m_E^{(45)}m_E^{(120)}}{m_E^{(45)^2}\left(1+\frac{1}{36}y_{45}^2\right)-16m_E^{(120)^2}\left(1+\frac{1}{384}y_{120}^2\right)},
\end{eqnarray}
and $\tau_i$ are listed in Appendix F.
\subsection{Top Quark Mass}
Using the computations given in Eq.(\ref{e3}) of Appendix E we display the mass matrix
for the top quark. We have
\newpage
\begin{eqnarray}\label{4.24}
\matrix{ {}^{(10_{16})}\!{{\overline{ {\Large {\textnormal
t}}}}}_{{ R\alpha}}&~~ {}^{(10_{45})}\!{\overline{ {\Large
{\textnormal t}}}}_{{ R\alpha}} &~~
{}^{({\overline{10}}_{45})}\!{\overline{ {\Large {\textnormal
t}}}}_{{ R\alpha}}& {}^{({{10}}_{120})}\!{\overline{ {\Large
{\textnormal t}}}}_{{ R\alpha}}&
{}^{({\overline{10}}_{120})}\!{\overline{ {\Large
{\textnormal t}}}}_{{ R\alpha}}}~~~\nonumber\\
 {\mathtt M}_{ { { t}}}~=~ \matrix{
{}^{(10_{16})}\!{{{ {\Large {\textnormal t}}}}}_{{ L}}^{\alpha}\cr
\cr {}^{({{10}}_{45})}\!{{{ {\Large {\textnormal t}}}}}_{{
L}}^{\alpha}\cr \cr {}^{({\overline{10}}_{45})}\!{{{ {\Large
{\textnormal t}}}}}_{{ L}}^{\alpha}\cr\cr {}^{({{10}}_{120})}\!{{{
{\Large {\textnormal t}}}}}_{{ L}}^{\alpha}\cr\cr {}^{({\overline
{10}}_{120})}\!{{{ {\Large {\textnormal t}}}}}_{{
L}}^{\alpha}}\left(\matrix{
 0 &{m_{{{ { t}}}}}^{(45)} & -m_U^{(45)} &{m_{{ { {t}}}}}^{(120)}& -m_U^{(120)}\cr\cr
{ m_{{ { { t}}}}}^{(45)}& 0 & -2m_F^{(45)}& 0&0 \cr\cr
 4m_U^{(45)} & -2m_F^{(45)} & 0& 0&0\cr\cr
{m_{{ { {t}}}}}^{(120)}&0&0&0&-2m_F^{(120)}\cr\cr
 4m_U^{(120)} &0&0 &-2m_F^{(120)} & 0}
 \right),
  \end{eqnarray}
  where
\begin{eqnarray}\label{4.25}
&{m_{{ {t}}}}^{(45)} =4f^{(120)}_{33}
\left[\frac{1}{\sqrt{10}}<{\bf
P}^5>+\frac{1}{\sqrt{6}}<\widetilde{\bf P}^5>\right],~~~~~{m_{{
{t}}}}^{(120)} =-\frac{8}{\sqrt{3}}f^{(120)}_{33}<{\bf
Q}^5>,&\nonumber\\
&m_U^{(120)}=\frac{16}{\sqrt{3}}f^{(120)}_{33}
q,~~~~~m_U^{(45)}=2\sqrt 2f^{(45)}_{33} p.&
\end{eqnarray}
In the limit ${m_{{ {t}}}}^{(45)},~{m_{{ {t}}}}^{(120)}$ are small,
the approximate light eigenvalue for the three  cases ${\mathsf D_1}, {\mathsf D_2}, {\mathsf D_3}$ is
\begin{eqnarray}\label{4.26}
{m_{{ {t}}}}^2\equiv\lambda_{{ t}_1}^2\approx \left\{\matrix{
\epsilon_t\times
\left(f^{(120)}_{33}<{\bf Q}^5>\right)^2;~~~~{\mathsf
D}_1
\cr\cr
\delta_t\times
 \left\{\matrix{\left(f^{(45)}_{33}<{\bf{
P}}^{\prime 5}>\left[\sqrt{3}\cos\vartheta_{\mathsf
D}+\sqrt{5}\sin\vartheta_{\mathsf D}\right]\right)^2;&{\mathsf
D}_2
\cr
\left(f^{(45)}_{33}<{\bf{\widetilde P}}^{\prime
5}>\left[-\sqrt{3}\sin\vartheta_{\mathsf
D}+\sqrt{5}\cos\vartheta_{\mathsf D}\right]\right)^2;&{\mathsf
D}_3,
}\right.}\right.
\end{eqnarray}
where $\epsilon_t$ and $\delta_t$ can are given in Appendix G.
The rotated fields can now be obtained from
\begin{eqnarray}\label{4.27}
\left(\matrix{\overline{{\Large {\textnormal t}}}_{{1 R}\alpha}\cr
\overline{{\Large {\textnormal t}}}_{{2 R}\alpha}\cr
\overline{{\Large {\textnormal t}}}_{{
3R}\alpha}\cr\overline{{\Large {\textnormal t}}}_{{ 4R}\alpha}
\cr\overline{{\Large {\textnormal t}}}_{{ 5R}\alpha}}\right)=V_{{
{t}}}\left(\matrix{{}^{({{10}}_{16})}\!\overline{{\Large
{\textnormal t}}}_{{ R}\alpha}\cr {}^{({
{10}}_{45})}\!\overline{{\Large {\textnormal t}}}_{{ R}\alpha}\cr
{}^{({\overline{10}}_{45})}\!\overline{{\Large {\textnormal
t}}}_{{ R}\alpha}\cr {}^{({{10}}_{120})}\!\overline{{\Large
{\textnormal t}}}_{{ R}\alpha} \cr
{}^{({\overline{10}}_{120})}\!\overline{{\Large {\textnormal
t}}}_{{ R}\alpha}}\right),~~~~~\left(\matrix{{\Large {\textnormal
t}}_{{ 1L}}^{\prime\alpha}\cr {\Large {\textnormal t}}_{{
2L}}^{\prime\alpha}\cr {\Large {\textnormal t}}_{{
3L}}^{\prime\alpha}\cr {\Large {\textnormal t}}_{{
4L}}^{\prime\alpha}\cr{\Large {\textnormal t}}_{{
5L}}^{\prime\alpha} }\right)=U_{{
{t}}}\left(\matrix{{}^{(10_{16})}\!{\Large {\textnormal t}}_{{
L}}^{\alpha}\cr {}^{(10_{45})}\!{\Large {\textnormal t}}_{{
L}}^{\alpha}\cr {}^{({\overline{10}}_{45})}\!{\Large {\textnormal
t}}_{{ L}}^{\alpha} \cr {}^{(10_{120})}\!{\Large {\textnormal
t}}_{{ L}}^{\alpha}\cr {}^{({\overline{10}}_{120})}\!{\Large
{\textnormal t}}_{{ L}}^{\alpha}} \right).
\end{eqnarray}

\noindent The mass terms in the Lagrangian are then given by
\begin{equation}\label{4.28}
\lambda_{{ t}_1}~ \overline{{\Large {\textnormal t}}}_{{1
R}\alpha} {\Large {\textnormal t}}_{{1
L}}^{\prime\alpha}~+~\lambda_{{ t}_2}~ \overline{{\Large
{\textnormal t}}}_{{2 R}\alpha}{\Large {\textnormal t}}_{{
2L}}^{\prime\alpha} ~+~\lambda_{{ t}_3} ~\overline{{\Large
{\textnormal t}}}_{{3 R}\alpha} {\Large {\textnormal t}}_{{
3L}}^{\prime\alpha}~+~\lambda_{{ t}_4}~ \overline{{\Large
{\textnormal t}}}_{{ 4R}\alpha} {\Large {\textnormal t}}_{{
4L}}^{\prime\alpha}~+~\lambda_{{ t}_5}~ \overline{{\Large
{\textnormal t}}}_{{ 5R}\alpha} {\Large {\textnormal t}}_{{
5L}}^{\prime\alpha}.
\end{equation}
In this case the rotation matrices   $U_t$ and $V_t$ that enter  Eq.(\ref{4.27})
 are given by
\begin{eqnarray}\label{4.29}
U_{{ {t}}}=\left(\matrix{t_1&t_{3+}&t_{3-}&0&0\cr
t_2&t_{4+}&t_{4-}& 0&0 \cr 0&0&0&t_{5-}&t_{5+}\cr 1&1&1&0&0\cr
    0&0&0&1&1
}\right),~~~V_{{
{t}}}=\left(\matrix{t_1^{\prime}&0&0&t_{4-}^{\prime}&t_{4+}^{\prime}\cr
t_2^{\prime}&0&0& t_{5+}^{\prime}&t_{5-}^{\prime} \cr
0&t_{3-}^{\prime}&t_{3+}^{\prime}&0&0\cr 1&0&0&1&1\cr
    0&1&1&0&0
}\right)
\end{eqnarray}
where $t_i$ and $t_i'$ are given in Appendix F.

\section{{${\mathbf b}-{\mathbf{\Tau}}-{\mathbf t} $} Unification\label{5}}
In this section we will consider the  light  doublet case  ${\mathsf
D}_1$ and determine the corresponding $ { b}-\tau$ and  $b-t$ unification
 conditions.
We consider $b-\tau$ unification first and here the
 equality ${h_{{ {b}}}}=\alpha {h_{{ {\tau}}}}$ gives a relationship
 between $y_{45}$ and $y_{120}$ which are defined by Eq.(\ref{g2})
 in Appendix G. One finds the following relation between them
\begin{eqnarray}\label{5.1}
(36\alpha^2-1)y_{45}^2y_{120}^4+ 8(96\alpha^2-51)y_{45}^2y_{120}^2
+ 36(\alpha^2-1)y_{120}^4 + 1124(4\alpha^2-9) (y_{45}^2+y_{120}^2)=0.
\end{eqnarray}
\begin{figure}[t]
  \begin{center}
             \includegraphics[width=9cm,height=9cm]{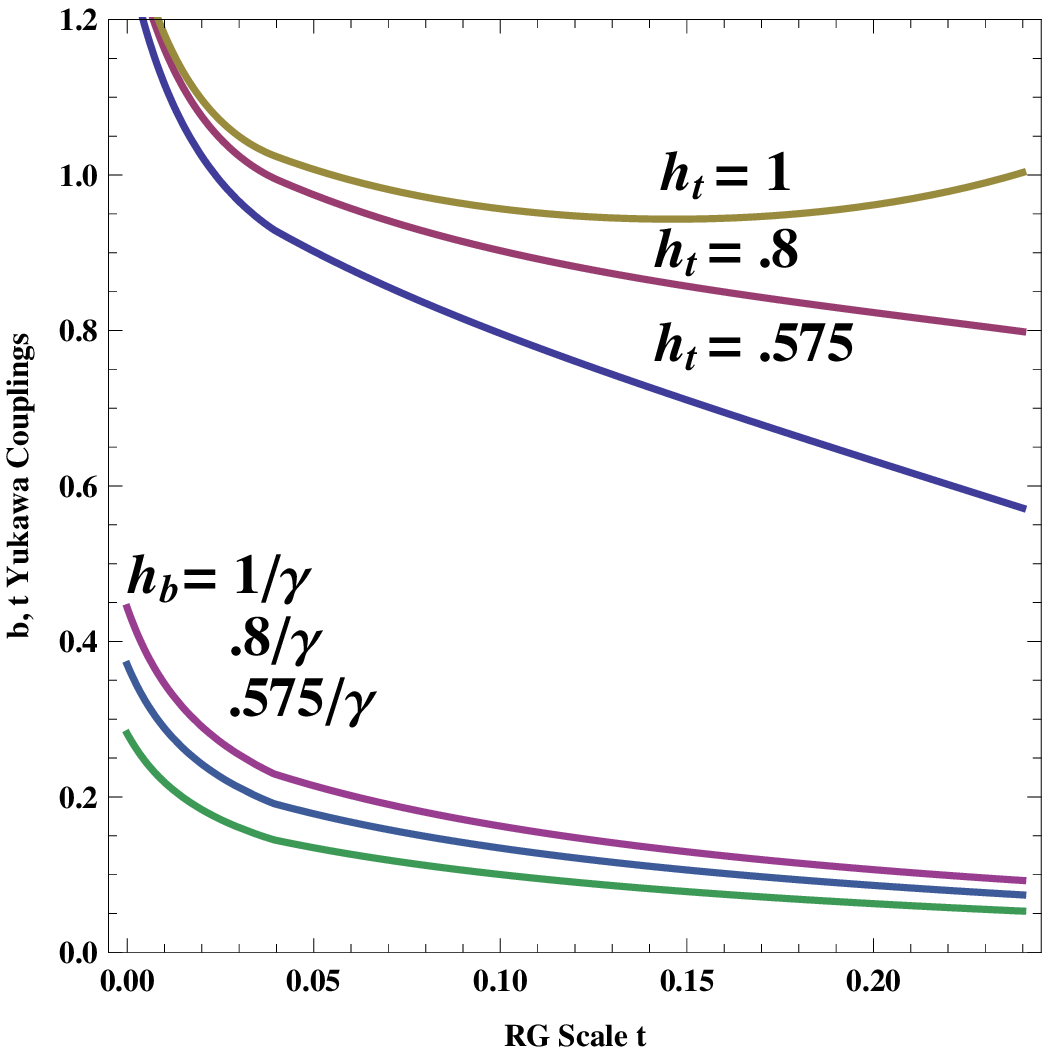}
              \includegraphics[width=9cm,height=9cm]{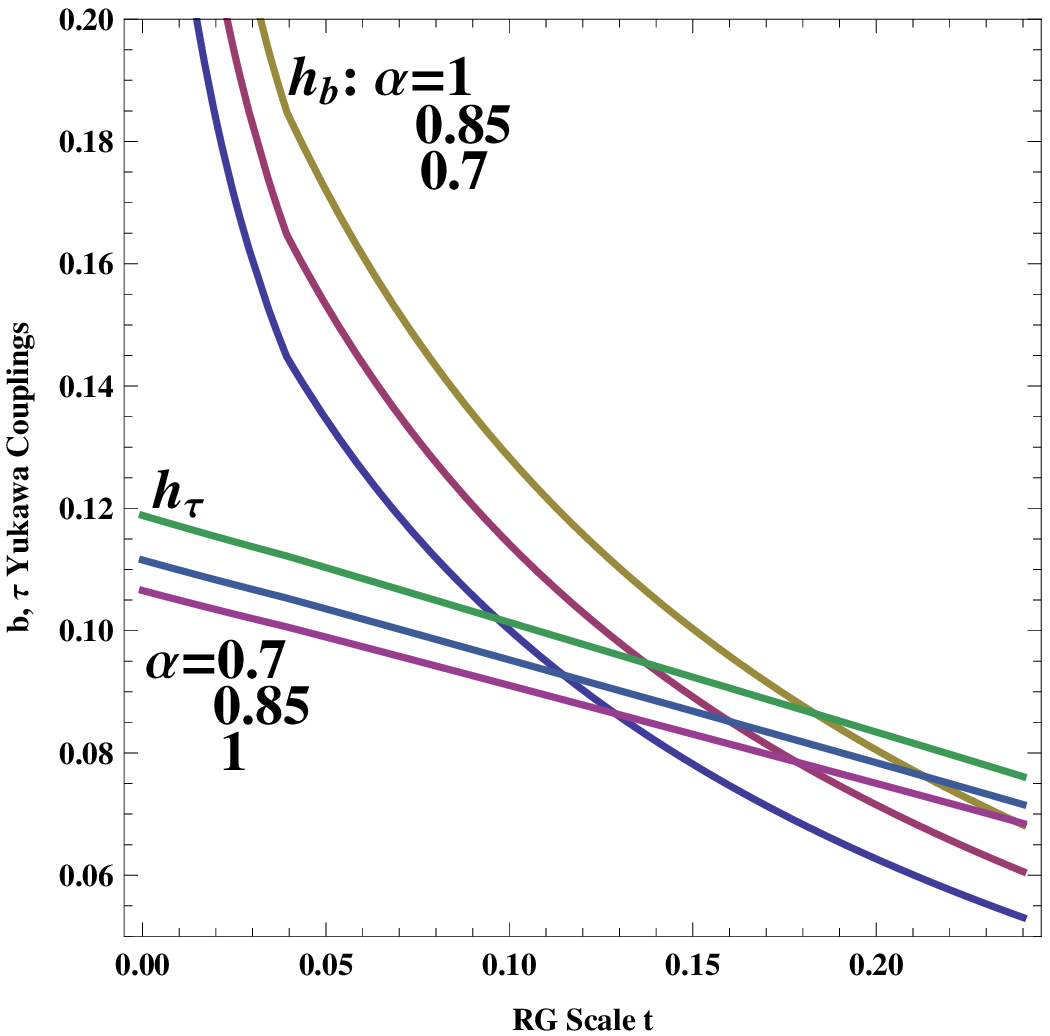}
                  \caption{Left panel: The RG evolution of the Yukawa coupling $h_b(Q)$  for the  bottom quark
                  and $h_{\tau}(Q)$ for the $\tau$ lepton
                  for $\alpha=1, 0.85, 0.7$ (see Table (2))  from the grand unification scale to
    the electroweak scale where t labeling the x-axis is defined so that $t=ln(Q({\rm GeV})/16\pi^2$.
    For the analysis of the figure in the left panel
    we have chosen $h_{t}(M_G)=0.575$ and $h_b(M_G)$ and $h_{\tau}(M_G)$ are then determined from Table (2).
    Right panel: The RG evolution of the Yukawa coupling $h_b(M_G)$
                  and of $h_{t}(M_G) $ for values of $h_t(M_G)=1, .8, .575$ and for $h_b(M_G)=1/\gamma,
                  .8/\gamma, .575/\gamma$
                    for the case $\alpha=.7$
                   from the grand unification scale to the electroweak scale where $\gamma=10.73$ as given in
                   Table (2). The convergence of the top
                   Yukawa couplings at low scales is
                       due to the presence of the Landau pole for the top Yukawa coupling.
                         The best fit to the $b,\tau,t$ masses given in
    Eq.(\ref{5.8}) occurs for $\alpha=.7$, $h_t(M_G)=0.575$.}
\label{bttfig}
  \end{center}
\end{figure}
Next we look at the ratio $m_t/m_b$, which we write in the following form
\beqn\label{5.11}
\frac{m_t}{m_b}=\frac{h_t}{h_b} \tan\beta,
~~ \tan\beta=\frac{<{\bf{ Q}}^{5}>}{<{\bf{P}}_5>}.
\eeqn
Our analysis using Eq.(\ref{4.14}) and Eq.(\ref{4.26})
gives for $\gamma=h_t/h_b$ at the GUT scale the result
\beqn\label{5.12}
\gamma=\frac{h_t}{h_b}= \frac{160}{3} \frac{f_{33}^{(120}}{f_{33}^{(45)}}
\frac{y_{45}}{y_{120}}
\sqrt{\frac{1+\frac{3}{32}y_{120}^2}{64+6\left(\frac{y_{120}}{y_{45}}\right)^2 +\frac{3}{8}y_{120}^2}}.
\eeqn
Since the values of $y_{45}$ and $y_{120}$ are constrained by Eq.(\ref{5.1}), we solve Eqs.(\ref{5.1}) and
(\ref{5.12}) to
determine the values of $h_t/h_b$ at the GUT scale as a function of $\alpha$. These results are
exhibited in Table(2).

\begin{center} \begin{tabular}{|c|c|c|c|c|c|c|c|}
\multicolumn{8}{c}{Table 2: Values of $\displaystyle{{h_{{ {t}}}}}/{{h_{{ {b}}}}}$ vs
$\displaystyle {{h_{{ {b}}}}}/{{h_{{ {\tau}}}}}$ at $Q=M_G$
}\cr \hline
 \hline\hline
$\alpha=h_b(M_G)/h_{\tau}(M_G)$ &  .7 & .75 & .8 & .85 & .9 & .95 &1\cr
\hline
$\gamma= h_t(M_G)/h_b(M_G)$ &  10.73 & 10.22 & 9.76& 9.40  & 8.99 & 8.65 & 8.35 \cr
\hline
\end{tabular}
\end{center}
\vskip 1cm
Eventually at low scales after the  VEV formation, the Higgs doublets
$({\bf{ Q}}^{5}, {\bf{P}}_5)$  will develop VEVs and lead to mass generation for the
top quark, for  the  bottom quark and for the tau lepton so that
\beqn\label{5.13}
m_{t}=\frac{h_{t} v \sin\beta}{\sqrt 2}, m_{b}=\frac{h_{b} v \cos\beta}{\sqrt 2},
m_{\tau}=\frac{h_{\tau} v \cos\beta}{\sqrt 2},
\eeqn
where
$ v=\sqrt{<{\bf Q}^{5}>^2+<{\bf P}_5>^2}$
and where numerically  $v=246$ GeV.

In a typical $SO(10)$ grand unification group where the electroweak symmetry is broken with
a $10$ plet of Higgs, a $b-\tau-t$ unification typically requires a large $\tan\beta$, i.e.,
a value of $\tan\beta$ as large as 40-50\cite{Ananthanarayan:1991xp}. However, from the analysis
above we find that a $b-\tau-t$ unification can occur for much smaller values of $\tan\beta$
(this was also noted in \cite{Babu:2006rp}).
 As is conventionally done, we  allow for the   possibility of Planck scale type corrections, and
 do not assume a rigid equality of the bottom and tau
   Yukawa couplings\cite{Barger:1992ac,Dasgupta:1995js,Baer:1999mc}.
   Specifically we  assume $h_{b}(M_G)=\alpha h_{\tau}(M_G)$ where $\alpha$ can deviate from unity.
With a given choice of $\alpha$, the ratio $h_t/h_b$ is then determined.
In Table (2) we list the allowed values of $h_t/h_b$  for $\alpha$ in the
range $.7-1.0$.
Using these boundary conditions we have carried out a one loop analysis of  the coupled
partial differential equations for the three Yukawa couplings $h_{\tau}, h_b, h_t$ and for the
three gauge coupling constants $g_3, g_2, g_1$ corresponding to  the three gauge groups $SU(3)_C, SU(2)_L, U(1)$
starting from the grand unification scale  $M_G$ and moving down to the electroweak scale.
In addition to the GUT boundary conditions on the Yukawa couplings $h_b, h_{\tau}, h_t$ given by
Table (2), the assumed range of $h_t$ is taken so that $|h_t|\leq 1$. Further, we assume that
 we have the GUT boundary conditions on the gauge couplings so that $g_3(M_G)=g_2(M_G)
=g_1(M_G) =g_G\simeq 0.7$.

The details  of our numerical computation to determine $b,\tau, t$ mass are as follows:
First we  do the RG analysis
 in the region between the GUT scale and some average weak SUSY scale
$M_S$, i.e., $M_S\leq Q\leq M_G$ using the boundary conditions at the GUT scale as discussed above.
Here we  use the RG evolution equations valid for MSSM.
Below the scale
$M_S$ we use the RG evolution equations appropriate for the Standard Model with proper matching done
at the scale $M_S$.
Numerical results for the evolution of the $b, \tau$ Yukawa couplings
$h_b(Q), h_{\tau}(Q)$ are plotted in the right panel of
 Fig.(1) and for $h_b(Q), h_t(Q)$ are plotted in the left  panel of
 Fig.(1) as a function of the scale $t=lnQ/16\pi^2$. The rapid convergence of the
 yukawas couplings for the top at low scale is due to the well known Landau pole singularity
 in the top Yukawa coupling\cite{Barger:1992ac,Nath:1995eq}.
In the RG analysis we compute the values of $m_t(m_t)$ which is related to the pole
mass $m_t^p$ by the relation
\beqn
m_t^p= m_t(m_t) [1+ \frac{4}{3\pi} \alpha_s(m_t)].
\eeqn
For the computation of $m_b(m_b)$ we use three  loop QCD and one loop EM evolution between the
scale $M_Z$ and the scale $m_b$. The procedure of the analysis is as  given in
\cite{Barger:1992ac,Dasgupta:1995js}.
 Using the result of Table 1 for $h_b/h_{\tau}=.7$ we find
a determination of $\tan\beta$ and  the following fit to the $b, t$ pole masses using the $\tau$ lepton mass
as an input $m_{\tau}=1.776$ GeV, so that
\beqn
\tan\beta=8,
~m_b= 4.53 ~{\rm GeV}, ~m_t^p=172.9~{\rm GeV}.~
\label{5.8}
\eeqn
The above are to be  compared with the experimental data on the $b$ quark and on the  top quark mass, i.e.,
$m_b=4.2^{+0.17}_{-0.07}$ GeV, $m_t=171.3\pm 1.1\pm 1.2$ GeV\cite{Amsler:2008zzb}.
Thus our one loop analysis is within
about 2$\sigma$ of the $b$ quark mass and within one sigma of the top quark mass.
We note that for the model above $b-\tau-t$ unification is achieved for a value which is
 much smaller than what typically appears in $b-\tau-t$ unification analyses.

\section{Tau Neutrino Mass\label{6}}
We begin by collecting all the relevant terms in the cubic Lagrangian that would
 contribute to neutrino mass:
\begin{eqnarray}\label{6.1}
\sum L = -2:~~~~{\mathsf
L}_{1,{\nu_{\tau}}}^{(120)}&=&-\frac{16}{\sqrt{15}}
f^{(120)}_{33}<{\bf Q}_5>{}^{(1_{16})}\!{{\overline{\Nu}}}
_{R\tau}{}^{({{5}}_{120})}\!{\Nu} _{L\tau}^{\mathsf c}+\textnormal{H.c.}, \nonumber\\
\sum L = 0:~~~~{\mathsf L}_{2,{\nu_{\tau}}}^{(120)}&=&16\sqrt{3}
f^{(120)}_{33}<{\bf Q}^5>{}^{(1_{16})}\!{{\overline{\Nu}}}
_{R\tau}{}^{({\overline{5}}_{120})}\!{\Nu} _{L\tau}+\textnormal{H.c.}, \nonumber\\
\sum L = 0:~~~~{\mathsf L}_{3,{\nu_{\tau}}}^{(120)}&=&8\sqrt{3}
f^{(120)}_{33}q~~{}^{(5_{120})}\!{{\overline{\Nu}}}
_{R\tau}{}^{({\overline{5}}_{16})}\!{\Nu} _{L\tau}+\textnormal{H.c.}, \nonumber\\
\sum L = 0:~~~~{\mathsf
L}_{4,{\nu_{\tau}}}^{(120)}&=&2m_F^{(120)}~~{}^{(5_{120})}\!{{\overline{\Nu}}}
_{R\tau}{}^{({\overline{5}}_{120})}\!{\Nu} _{L\tau}+\textnormal{H.c.}, \nonumber\\
\sum L = -2:~~~~{\mathsf
L}_{6,{\nu_{\tau}}}^{(45)}&=&m_F^{(45)}~~{}^{(1_{45})}\!{{\overline{\Nu}}}
_{R\tau}{}^{({{1}}_{45})}\!{\Nu} _{L\tau}^{\mathsf
c}+\textnormal{H.c.}, \nonumber\\
\sum L = -2:~~~~{\mathsf
L}_{{\rm quartic},{\nu_{\tau}}}&=&m_F^{(16)}~~{}^{(1_{16})}\!{{\overline{\Nu}}}
_{R\tau}{}^{({{1}}_{16})}\!{\Nu} _{L\tau}^{\mathsf
c}+\textnormal{H.c.},
\end{eqnarray}
where we have identified lepton number ($L$) conserving
(Dirac-like)  and lepton number violating
violating (Majorana-like) interactions. There are no couplings of the $1_{45}$ state with
the other neutrino states
\footnote{
The coupling  ${\overline{5}}_{16}\cdot1_{45}\cdot5_{\overline{144}}$ is
disallowed by the $U(1)$ quantum number constraint and thus does not appear
among the terms generating a Dirac mass for the neutrino in Eq.(\ref{6.1}).},
and thus the $1_{45}$ which is superheavy can be integrated out and does not
enter into the rest of the analysis.   Further, the states $5_{120}$ and $\bar 5_{120}$
are also superheavy and can be eliminated leaving an effective Dirac mass term
involving $1_{16}$ and $\bar 5_{16}$ and a Majorana mass term for the $1_{16}$ which
is the last term in Eq.(\ref{6.1}). These remaining interactions then produce a
Type I
See Saw
mass for the light tau neutrino which is given by
\beqn\label{6.11}
m_{\nu_{\tau}} \simeq \frac{m_D^2\sin^2\beta}{m_{F}^{(16)}},
~~m_D\simeq 96 (f_{33}^{(120)})^2 v  \left(\frac{q}{m_F^{(120)}}\right)
\eeqn
It is interesting to note  the dependence of the neutrino mass on $\sin\beta$ in Eq.(\ref{6.11})
which reflects
the chose connection of the neutrino mass with electroweak physics.
Now the choice $f_{33}^{(120)}\sim .2$, $q/M_{F}^{(120}\sim 1$ and $M_F^{(16)}=10^{16}$ GeV,
and $v=246$ GeV, $\sin\beta \sim 1$ leads to a Type I See Saw mass of
 $m_{\nu_{\tau}} \simeq  0.1 eV$.
A more complete analysis of fermion masses including also the Type II and Type III
See Saw masses in the unified Higgs framework will be given elsewhere\cite{bgns3}.

\section{Proton stability}\label{7}
Grand unified models of  particle interactions typically lead to higgsino mediated proton decay
via baryon and lepton ($B\&L$) number violating dimension five operators. There are severe
restrictions on the size of such operators from the current limits on proton decay\cite{Nath:2006ut}
(For some recent works on  proton decay see\cite{pdecay,Nath:2007eg} and for
 a recent review of experiment  see \cite{Rubbia:2009md}).
 In the present context such $B\&L$
 violating dimension five operators arise from the quartic interactions of the type $16\cdot16\cdot144\cdot144$
and $16\cdot16\cdot\overline{144}\cdot\overline{144}$ operators. In addition there are $B\&L$
 violating dimension five
operators from the $16\cdot120\cdot144$ and from $16\cdot45\cdot\overline{144}$.  In Ref.\cite{Nath:2007eg}
a cancelation mechanism to suppress proton decay was advocated and some specific examples
of such a cancelation were also exhibited there. Such a cancelation mechanism could also
work for dimension five operators arising from the above mentioned interactions.
An analysis of proton lifetime within the unified Higgs framework without invoking
the cancellation mechanism is given in a recent work\cite{Wu:2009zzh} where the
decay lifetime $\tau(p\to \bar \nu K^+)$ is found to lie in the range
$(1.4\times 10^{35}-1.9\times 10^{33})$yr for $\tan\beta$ in the range
$\tan\beta=2-20$. This puts the proton lifetime in the unified Higgs model at
the edge of detection in improved proton decay experiment\cite{Rubbia:2009md,Raby:2008pd}.

\section{Conclusion\label{8}}
In this work we have carried out an analysis of the Yukawa couplings that arise  from
quartic couplings of the type
$\frac{1}{\Lambda}16\cdot16\cdot144\cdot144$ which can contribute to the quark-charged lepton and neutrino masses.
Such Yukawas arise
after  spontaneous breaking of the $SO(10)$ gauge symmetry at the grand unification scale
where the $144$ and $\overline {144}$ of the Higgs multiplets develop vacuum expectation
values. Computations of these Yukawa couplings are  non-trivial  involving special techniques
which are utilized for their computation in this work. Thus after the $144$  of the Higgs fields
develop superheavy VEVs, the interaction $\frac{1}{\Lambda}16\cdot16\cdot144\cdot144$ generates Yukawa couplings
for the quarks, charged leptons and the neutrinos which we calculate. These Yukawas are typically
$O(\frac{<144>}{\Lambda})\sim O(\frac{M_G}{M_{St}})$ and thus typically of size $ 10^{-2}$. These
Yukawa couplings are thus appropriate for the analysis of the mass matrices for the first and second
 generation masses.  However, these Yukawas are too small for the generation of third
generation fermion masses. To allow for significantly larger third generation fermion masses
we considered cubic couplings involving $45$ and $120$ plets of matter fields, i.e., couplings
of the type $16\cdot45\cdot\overline{144}$ and $16\cdot120\cdot144$. These interactions lead to cubic couplings
of size  appropriate for the third generation. An analysis for the quarks,  charged lepton and neutrino
masses was also given. It is shown that the $144$ plet Higgs couplings allow for
a See Saw mechanism for the generation of neutrino masses. The analysis of the Yukawa interactions
given here would be of significant value in the further development of the $SO(10)$ phenomenology
in the unified Higgs framework.

\noindent
{\bf Acknowledgements:}
 We thank K.S. Babu and  Ilia Gogoladze for discussions. This  work was supported in part by NSF grant PHY-0757959.
\clearpage

\begin{center}
    \begin{tabular}{||c||l||c||l|}
\multicolumn{4}{c}{Table 1: Fermion mass parameters from quartic couplings}\cr
    \hline
 $A^{(10)}_{1~\acute {a}\acute
 {b}}$&$-48\zeta_{\acute{a}\acute{b},\acute{c}\acute{d}}^{^{(10)(+)}}q_{\acute{c }}<{{\bf
Q}}_{{\acute {d}}}^{ 5}>$ & $A^{(10)}_{2~\acute {a}\acute
{b}}$&$\frac{24}{\sqrt
5}\xi_{\acute{a}\acute{b},\acute{c}\acute{d}}^{^{(10)(+)}}p_{\acute{c
}}<{{\bf
P}}_{{\acute {d}}}^{\prime 5}>\cos\vartheta_{\mathsf D}$\\
\hline

$A^{(10)}_{3~\acute {a}\acute {b}}$&$\frac{24}{\sqrt
5}\xi_{\acute{a}\acute{b},\acute{c}\acute{d}}^{^{(10)(+)}}p_{\acute{c
}}<{\widetilde{\bf P}}_{{\acute {d}}}^{\prime
5}>\sin\vartheta_{\mathsf D}$ &$A^{(10)}_{4~\acute {a}\acute
{b}}$&$-40\sqrt
3\xi_{\acute{a}\acute{b},\acute{c}\acute{d}}^{^{(10)(+)}}p_{\acute{c
}}<{{\bf P}}_{{\acute {d}}}^{\prime 5}>\sin\vartheta_{\mathsf
D}$\\
\hline

$A^{(10)}_{5~\acute {a}\acute {b}}$&$40\sqrt
3\xi_{\acute{a}\acute{b},\acute{c}\acute{d}}^{^{(10)(+)}}p_{{\acute{c}}}<{\widetilde{\bf
P}}_{{\acute {d}}}^{\prime 5}>\cos\vartheta_{\mathsf D}$&$
A^{(10)}_{6~\acute {a}\acute
 {b}}$&$6\lambda_{\acute{a}\acute{d},\acute{b}\acute{c}}^{^{(10)}}q_{\acute{c }}<{{\bf
Q}}_{{\acute {d}}}^{ 5}>$\\
\hline

 $ A^{(45)}_{1~\acute {a}\acute {b}}$&$-\frac{2}{\sqrt
5}(4\lambda_{{\acute {a}}{\acute {c}},{\acute {b}}{\acute
{d}}}^{^{(45)}}-5\lambda_{{\acute {a}}{\acute {d}},{\acute
{b}}{\acute {c}}}^{^{(45)}})p_{\acute{c }}<{\bf {
P}}_{\acute{d}}^{\prime 5}>\cos\vartheta_{\mathsf D}$&$
 A^{(45)}_{2~\acute
{a}\acute {b}}$&$-\frac{2}{\sqrt 5}(4\lambda_{{\acute {a}}{\acute
{c}},{\acute {b}}{\acute {d}}}^{^{(45)}}-5\lambda_{{\acute
{a}}{\acute {d}},{\acute {b}}{\acute {c}}}^{^{(45)}})p_{\acute{c
}}<{\widetilde{\bf { P}}}_{\acute{d}}^{\prime
5}>\sin\vartheta_{\mathsf D}$\\
\hline

 $A^{(45)}_{3~\acute {a}\acute
{b}}$&$\frac{2}{\sqrt 3}(2\lambda_{{\acute {a}}{\acute
{c}},{\acute {b}}{\acute {d}}}^{^{(45)}}-9\lambda_{{\acute
{a}}{\acute {d}},{\acute {b}}{\acute {c}}}^{^{(45)}})p_{\acute{c
}}<{{\bf { P}}}_{\acute{d}}^{\prime 5}>\sin\vartheta_{\mathsf D}$&
 $A^{(45)}_{4~\acute
{a}\acute {b}}$&$-\frac{2}{\sqrt 3}(2\lambda_{{\acute {a}}{\acute
{c}},{\acute {b}}{\acute {d}}}^{^{(45)}}-9\lambda_{{\acute
{a}}{\acute {d}},{\acute {b}}{\acute {c}}}^{^{(45)}})p_{\acute{c
}}<{\widetilde{\bf { P}}}_{\acute{d}}^{\prime
5}>\cos\vartheta_{\mathsf D}$\\
\hline

 $A^{(54)}_{1~\acute {a}\acute {b}}$&$-\frac{4}{\sqrt
3}\lambda_{{\acute {a}}{\acute {c}},{\acute {b}}{\acute
{d}}}^{^{(54)}}p_{\acute{c }}<{{\bf { P}}}_{\acute{d}}^{\prime
5}>\sin\vartheta_{\mathsf D}$&$
 A^{(54)}_{2~\acute
{a}\acute {b}}$&$\frac{4}{\sqrt 3}\lambda_{{\acute {a}}{\acute
{c}},{\acute {b}}{\acute {d}}}^{^{(54)}}p_{\acute{c
}}<{\widetilde{\bf { P}}}_{\acute{d}}^{\prime
5}>\cos\vartheta_{\mathsf D}$\\
\hline

 $A^{(54)}_{3~\acute {a}\acute
{b}}$&$\frac{3}{\sqrt 5}\lambda_{{\acute {a}}{\acute {c}},{\acute
{b}}{\acute {d}}}^{^{(54)}}p_{\acute{c }}<{{\bf {
P}}}_{\acute{d}}^{\prime 5}>\cos\vartheta_{\mathsf D}$& $
A^{(54)}_{4~\acute {a}\acute {b}}$&$\frac{3}{\sqrt
5}\lambda_{{\acute {a}}{\acute {c}},{\acute {b}}{\acute
{d}}}^{^{(54)}}p_{\acute{c }}<{\widetilde{\bf {
P}}}_{\acute{d}}^{\prime 5}>\sin\vartheta_{\mathsf D}$\\
\hline

$ A^{(126)}_{~\acute {a}\acute
{b}}$&$-\frac{32}{5}\varrho_{\acute{a}\acute{b},\acute{c}\acute{d}}^{^{(126,\overline{126})(+)}}q_{\acute{c
}}<{{\bf Q}}_{{\acute {d}}}^{5}> $&&\\
\hline

&&&\\
\hline

$B^{(10)}_{1~\acute {a}\acute {b}}$&$\frac{24}{\sqrt
5}\zeta_{\acute{a}\acute{b},\acute{c}\acute{d}}^{^{(10)(+)}}q_{\acute{c
}}<{\bf { Q}}_{\acute{d}5}^{\prime}>\cos\vartheta_{\mathsf D}$&$
 B^{(10)}_{2~\acute
{a}\acute {b}}$&$\frac{24}{\sqrt
5}\zeta_{\acute{a}\acute{b},\acute{c}\acute{d}}^{^{(10)(+)}}q_{\acute{c
}}<{\widetilde{\bf { Q}}}_{\acute{d}5}^{\prime}>\sin\vartheta_{\mathsf D}$\\
\hline

$B^{(10)}_{3~\acute {a}\acute {b}}$&$-40\sqrt
3\zeta_{\acute{a}\acute{b},\acute{c}\acute{d}}^{^{(10)(+)}}q_{\acute{c
}}<{\bf { Q}}_{\acute{d}5}^{\prime}>\sin\vartheta_{\mathsf D}$&$
B^{(10)}_{4~\acute {a}\acute {b}}$&$40\sqrt
3\zeta_{\acute{a}\acute{b},\acute{c}\acute{d}}^{^{(10)(+)}}q_{\acute{c
}}<{\widetilde{\bf { Q}}}_{\acute{d}5}^{\prime}>\cos\vartheta_{\mathsf D}$\\
\hline

$ B^{(10)}_{5~\acute {a}\acute
{b}}$&$48\xi_{\acute{a}\acute{b},\acute{c}\acute{d}}^{^{(10)(+)}}p_{\acute{c
}}<{\bf { P}}_{\acute{d}5}>$&$ B^{(10)}_{6~\acute {a}\acute
{b}}$&$-\frac{2}{\sqrt
5}\lambda_{\acute{a}\acute{c},\acute{b}\acute{d}}^{^{(10)}}q_{\acute{c
}}<{\bf { Q}}_{\acute{d}5}^{\prime}>\cos\vartheta_{\mathsf D}$\\
\hline

$ B^{(10)}_{7~\acute {a}\acute {b}}$&$-\frac{2}{\sqrt
5}\lambda_{\acute{a}\acute{c},\acute{b}\acute{d}}^{^{(10)}}q_{\acute{c
}}<{\widetilde{\bf {
Q}}}_{\acute{d}5}^{\prime}>\sin\vartheta_{\mathsf D}$&$
B^{(45)}_{~\acute {a}\acute
{b}}$&$-2\lambda_{\acute{a}\acute{d},\acute{b}\acute{c}}^{^{(45)}}p_{\acute{c
}}<{\bf { P}}_{\acute{d}5}>$\\
\hline

$ B^{(54)}_{~\acute {a}\acute
{b}}$&$-10\lambda_{\acute{a}\acute{d},\acute{b}\acute{c}}^{^{(54)}}p_{\acute{c
}}<{\bf { P}}_{\acute{d}5}>$&$B^{(126)}_{1~\acute {a}\acute
{b}}$&$\frac{11}{15\sqrt
5}\varrho_{\acute{a}\acute{b},\acute{c}\acute{d}}^{^{(126,\overline{126})(+)}}q_{\acute{c
}}<{{\bf Q}}_{{\acute {d}}5}^{\prime}>\cos\vartheta_{\mathsf D}$\\
\hline

$ B^{(126)}_{2~\acute {a}\acute {b}}$&$\frac{11}{15\sqrt
5}\varrho_{\acute{a}\acute{b},\acute{c}\acute{d}}^{^{(126,\overline{126})(+)}}q_{\acute{c
}}<{\widetilde{\bf {
Q}}}_{\acute{d}5}^{\prime}>\sin\vartheta_{\mathsf D}$&$
B^{(126)}_{3~\acute {a}\acute {b}}$&$-\frac{7}{15\sqrt
3}\varrho_{\acute{a}\acute{b},\acute{c}\acute{d}}^{^{(126,\overline{126})(+)}}q_{\acute{c
}}<{{\bf Q}}_{{\acute {d}}5}^{\prime}>\sin\vartheta_{\mathsf D}$\\
\hline

$ B^{(126)}_{4~\acute {a}\acute {b}}$&$\frac{7}{15\sqrt
3}\varrho_{\acute{a}\acute{b},\acute{c}\acute{d}}^{^{(126,\overline{126})(+)}}q_{\acute{c
}}<{\widetilde{\bf {
Q}}}_{\acute{d}5}^{\prime}>\cos\vartheta_{\mathsf D} $&&\\
\hline

&&&\\
\hline

$C^{(45)}_{~\acute {a}\acute
{b}}$&$30\lambda_{\acute{a}\acute{c},\acute{b}\acute{d}}^{^{(45)}}p_{\acute{c
}}p_{\acute{d }}$&$ C^{(54)}_{~\acute {a}\acute
{b}}$&$-30\lambda_{\acute{a}\acute{c},\acute{b}\acute{d}}^{^{(54)}}p_{\acute{c
}}p_{\acute{d }}$\\
\hline

 $C^{(10)}_{1~\acute {a}\acute {b}}$&$-\frac{4}{\sqrt
5}\lambda_{\acute{a}\acute{c},\acute{b}\acute{d}}^{^{(10)}} <{{\bf
Q}}_{{\acute {c}}}^{ 5}><{{\bf Q}}_{{\acute {d}5}}^{\prime
}>\cos\vartheta_{\mathsf D}$&$ C^{(10)}_{2~\acute {a}\acute
{b}}$&$-\frac{4}{\sqrt
5}\lambda_{\acute{a}\acute{c},\acute{b}\acute{d}}^{^{(10)}} <{{\bf
Q}}_{{\acute {c}}}^{ 5}><{\widetilde{\bf Q}}_{{\acute
{d}5}}^{\prime }>\sin\vartheta_{\mathsf D}$\\
\hline

$ C^{(126)}_{1~\acute {a}\acute {b}}$&$-\frac{16}{15\sqrt
5}\varrho_{\acute{a}\acute{b},\acute{c}\acute{d}}^{^{(126,\overline{126})(+)}}
<{{\bf Q}}_{{\acute {c}}}^{ 5}><{{\bf Q}}_{{\acute {d}5}}^{\prime
}>\cos\vartheta_{\mathsf D}$&$ C^{(126)}_{2~\acute {a}\acute
{b}}$&$-\frac{16}{15\sqrt
5}\varrho_{\acute{a}\acute{b},\acute{c}\acute{d}}^{^{(126,\overline{126})(+)}}
<{{\bf Q}}_{{\acute {c}}}^{ 5}><{\widetilde{\bf Q}}_{{\acute
{d}5}}^{\prime }>\sin\vartheta_{\mathsf D}$\\
\hline

&&&\\
\hline

$D^{(45)}_{1~\acute {a}\acute {b}}$&$-\frac{1}{
4}(\lambda_{{\acute {a}}{\acute {c}},{\acute {b}}{\acute
{d}}}^{^{(45)}}-5\lambda_{{\acute {a}}{\acute {d}},{\acute
{b}}{\acute {c}}}^{^{(45)}})<{\bf { P}}_{\acute{c}}^{\prime
5}><{\bf { P}}_{\acute{d}}^{\prime 5}>\cos^2\vartheta_{\mathsf
D}$&$ D^{(45)}_{2~\acute {a}\acute {b}}$&$-\frac{1}{
4}(\lambda_{{\acute {a}}{\acute {c}},{\acute {b}}{\acute
{d}}}^{^{(45)}}-5\lambda_{{\acute {a}}{\acute {d}},{\acute
{b}}{\acute {c}}}^{^{(45)}})<{\widetilde{\bf {
P}}}_{\acute{c}}^{\prime 5}><{\widetilde{\bf {
P}}}_{\acute{d}}^{\prime 5}>\sin^2\vartheta_{\mathsf D}$\\
\hline

$ D^{(45)}_{3~\acute {a}\acute {b}}$&$(\lambda_{{\acute
{a}}{\acute {c}},{\acute {b}}{\acute
{d}}}^{^{(45)}}+\lambda_{{\acute {a}}{\acute {d}},{\acute
{b}}{\acute {c}}}^{^{(45)}})<{{\bf { P}}}_{\acute{c}}^{\prime
5}><{\widetilde{\bf { P}}}_{\acute{d}}^{\prime
5}>\cos\vartheta_{\mathsf D}\sin\vartheta_{\mathsf D}$&$
D^{(54)}_{1~\acute {a}\acute {b}}$&$\frac{3}{
80}(67\lambda_{{\acute {a}}{\acute {c}},{\acute {b}}{\acute
{d}}}^{^{(54)}}-15\lambda_{{\acute {a}}{\acute {d}},{\acute
{b}}{\acute {c}}}^{^{(54)}})<{\bf { P}}_{\acute{c}}^{\prime
5}><{\bf { P}}_{\acute{d}}^{\prime
5}>\cos^2\vartheta_{\mathsf D}$\\
\hline

$ D^{(54)}_{2~\acute {a}\acute
{b}}$&$\frac{3}{80}(67\lambda_{{\acute {a}}{\acute {c}},{\acute
{b}}{\acute {d}}}^{^{(54)}}-15\lambda_{{\acute {a}}{\acute
{d}},{\acute {b}}{\acute {c}}}^{^{(54)}})<{\widetilde{\bf {
P}}}_{\acute{c}}^{\prime 5}><{\widetilde{\bf {
P}}}_{\acute{d}}^{\prime 5}>\sin^2\vartheta_{\mathsf D}$&$
D^{(54)}_{3~\acute {a}\acute {b}}$&$\frac{39}{20}(\lambda_{{\acute
{a}}{\acute {c}},{\acute {b}}{\acute
{d}}}^{^{(54)}}+\lambda_{{\acute {a}}{\acute {d}},{\acute
{b}}{\acute {c}}}^{^{(54)}})<{{\bf { P}}}_{\acute{c}}^{\prime
5}><{\widetilde{\bf { P}}}_{\acute{d}}^{\prime
5}>\cos\vartheta_{\mathsf D}\sin\vartheta_{\mathsf D}$\\
\hline
\end{tabular}
\end{center}
Table caption: Fermion  mass parameters arising
from the interactions of the
 $144+\overline{144}$ with the 16 plet of matter after spontaneous
 breaking at the electroweak scale.  All the coupling constants
 such as $\zeta_{\acute{a}\acute{b},\acute{c}\acute{d}}$
 are scaled by the inverse mass $\Lambda^{-1}$ (see Eq.(\ref{1.2})) which is implicit in the
 definition of these couplings.
\section{Appendix A: Formalism and Calculational Techniques\label{a}}

In this appendix we give a brief discussion of the formalism and of the calculational  techniques used
in the analysis of this paper.
In the analysis we need couplings of the 144-plet and $\overline{144}$ plets of Higgs
which includes  self couplings of the type $(144\cdot\overline{144})^2$,   quartic couplings
 with matter fields of the type $(16\cdot16)(144\cdot144)$, $(16\cdot16)(\overline {144}\cdot\overline {144})$,
$(16\cdot144)(16\cdot144)$,  $(16\cdot \overline{144})(16\cdot\overline{144}),$
and cubic couplings with the matter fields  $16\cdot120\cdot144$ and $16\cdot 45\cdot\overline{144}$.
 In $SO(10)$ the $144, \overline{144}$ are
constrained  vector-spinors which are gotten from  the vector spinor $160$ by imposition
of a  constraint which removes $16$ components. Thus the 144-dimensional
 vector spinor $|\Upsilon_{(\pm)\mu}>$ is defined by
  \beq \label{a1}
  \Gamma_{\mu}
|\Upsilon_{(\pm)\mu}> =0,
\eeq where $\Gamma_{\mu}$ satisfy a
rank-10 Clifford algebra
\begin{equation}\label{a2}
\{\Gamma_{\mu},\Gamma_{\nu}\}=2\delta_{\mu\nu},
\end{equation}
and where $\mu,\nu$ are the $SO(10)$ indices and take on the
values $1,..,10$. We introduce now the notation for the components in $144$ and
$\overline{144}$ in an $SU(5)\times U(1)$ decomposition.
For the $144$ we have
\beqn\label{a3}
144= 5 ({\bf{\widehat Q}}^i)+\bar 5 ({\bf{\widehat Q}}_i)+ 10 ({\bf{\widehat Q}}^{ij}) + 15 ({\bf{\widehat Q}}_{(S)}^{ij})+ 24({\bf{\widehat Q}}^i_j) +
40({\bf{\widehat Q}}^{ijk}_l) + \overline{45}({\bf{\widehat Q}}^{i}_{jk}),
\eeqn
where $i,j,k,..$ are the $SU(5)$ indices which take on the
values $1,..,5$. Similarly for $\overline{144}$ we introduce the notation
 \beqn  \label{a4}
 \overline{144} =\bar 5
({\bf{\widehat P}}_i)+5 ({\bf{\widehat P}}^i)+ \overline{10}({\bf{\widehat P}}_{ij}) + \overline{15}({\bf{\widehat P}}^{(S)}_{ij}) + 24
({\bf{\widehat P}}^i_j) + \overline{40}({\bf{\widehat P}}^l_{ijk}) +
45({\bf{\widehat P}}^{ij}_{k}).
\label{2.4}
\eeqn
Note that fields with a hat, $~\widehat{  }~$ stand for chiral superfields, while the
ones without a hat represents the charge scalar component
of the corresponding superfield.

For convenience of computations we will use  the oscillator expansion technique\cite{ms,wilczek,ns}.
(For an alternative technique see, e.g., \cite{Aulakh:2002zr}).
In the oscillator method one uses
for an $SO(10)$ group
 a set of 5 operators $b_i, b_i^{\dagger}$ (i=1...5)
which obey the anti-commutation relations
\beqn\label{a5}
\{b_i,b_j^{\dagger}\}=\delta_{ij}, ~~\{b_i,b_j\} =0=\{b_i^{\dagger}, b_j^{\dagger}\},
\eeqn
and the set of $10$ operators $\Gamma_{\mu}$  defined by
\beqn\label{a6}
\Gamma_{2i}=(b_i+b_i^{\dagger}), ~~\Gamma_{2i-1}=-i(b_i-b_i^{\dagger}).
\eeqn
Further,  in  the oscillator expansion\cite{ms,wilczek}
the $16 (|\Psi_{(+)}>)$ and
$\overline{16}(|\Psi_{(-)}>)$ have the form
\begin{equation}\label{a7}
|\Psi_{(+)}>=|0>{\bf
\widehat P}+\frac{1}{2}b_i^{\dagger}b_j^{\dagger}|0>{\bf
\widehat P}^{ij} +\frac{1}{24}\epsilon^{ijklm}b_j^{\dagger}
b_k^{\dagger}b_l^{\dagger}b_m^{\dagger}|0>{\bf P}_{i},
\end{equation}
\begin{equation}\label{a8}
|\Psi_{(-)}>=b_1^{\dagger}b_2^{\dagger}
b_3^{\dagger}b_4^{\dagger}b_5^{\dagger}|0>{\bf\widehat Q}
+\frac{1}{12}\epsilon^{ijklm}b_k^{\dagger}b_l^{\dagger}
b_m^{\dagger}|0>{\bf\widehat Q}_{ij}+b_i^{\dagger}|0>{\bf\widehat Q}^i.
\end{equation}
The 160 plet spinor ($|\Psi_{(+)\mu}>$) and the $\overline{160}$ plet spinor ($|\Psi_{(-)\mu}>$) are
defined by\cite{Babu:2005gx}\begin{equation}\label{a9}
|\Psi_{(+)\acute{a}\mu}>=|0>{\bf\widehat P}_{\acute{a}\mu}+\frac{1}{2}b_i^{\dagger}b_j^{\dagger}|0>{\bf\widehat P}_{\acute{a}\mu}^{ij} +\frac{1}{24}\epsilon^{ijklm}b_j^{\dagger}
b_k^{\dagger}b_l^{\dagger}b_m^{\dagger}|0>{\bf\widehat P}_{\acute{a}i\mu},
\end{equation}
\begin{equation}\label{a10}
|\Psi_{(-)\acute{b}\mu}>=b_1^{\dagger}b_2^{\dagger}
b_3^{\dagger}b_4^{\dagger}b_5^{\dagger}|0>{\bf\widehat Q}_{\acute{b}\mu}
+\frac{1}{12}\epsilon^{ijklm}b_k^{\dagger}b_l^{\dagger}
b_m^{\dagger}|0>{\bf\widehat Q}_{\acute{b}ij\mu}+b_i^{\dagger}|0>{\bf\widehat Q}_{\acute{b}\mu}^i.
\end{equation}
The $144$ and $\overline{144}$ spinors $|\Upsilon_{(\pm)\mu}>$ are  created from
 $|\Psi_{(\pm)\mu}>$ by removing the  $16$ components
  $\Gamma_{\mu}|\Psi_{(\pm)\mu}>$  from $|\Psi_{(\pm)\mu}>$.

\section{Appendix B: Details of analysis of the quartic couplings\label{b}}
In this Appendix we give further details  of the computation of the couplings that go
in the analysis of Sec.(\ref{3}).
First we note that the  interactions
$(16\cdot 16)_{10}(144\cdot 144)_{10}$, $(16\cdot 16)_{10}(\overline{144}\cdot \overline{144})_{10}$,
$(16\cdot 16)_{120}(144\cdot 144)_{120}$, $(16\cdot 16)_{120}(\overline{144}\cdot \overline{144})_{120}$,
and $(16\cdot 16)_{\overline{126}}(144\cdot 144)_{126}$
have already been computed in \cite{Nath:2005bx}.
For the case when there is only generation of the $144+\overline{144}$ of Higgs multiplets,
 the interactions $(16\cdot 16)_{120}(144\cdot 144)_{120}$
and $(16\cdot 16)_{120}(\overline{144}\cdot\overline{144})_{120}$
do not contribute. On the other hand, for the analysis  of Sec.(\ref{3}) we also need
the interactions $(16\cdot 144)_{10}(16\cdot 144)_{10}$,
$(16\cdot\overline{ 144})_{45}(16\cdot\overline{144})_{45}$,
$(16\cdot\overline{ 144})_{54}(16\cdot\overline{144})_{54}$.
An explicit computations of these has not been given before and so in this
Appendix we give an analysis of these interactions.
 We now give the details of the analysis.
\subsection{$\mathbf{(16\cdot 144)_{10}(16\cdot
144)_{10}}$ coupling}
The dimension-5 operator of the form $(16\cdot 144)_{10}(16\cdot
144)_{10}$ arises from the following superpotential
\begin{eqnarray}\label{b1}
{\mathsf W}^{^{(10)'}}=\frac{1}{2}\Phi_{\mu{\cal S}} {\cal
M}^{^{(10)}}_{{\cal S}{\cal S}'} \Phi_{\mu {\cal
S}'}+j_{{\acute{a}\acute{b}}}^{^{(10)}}<\Psi^*_{(+)\acute{a}}|B|\Upsilon_{(-)\acute{b}\mu}>k_{_{{
S}}}^{^{(10)}}\Phi_{\mu {\cal S}}.
\end{eqnarray}
Elimination of the  $10$ plet gives,
\begin{eqnarray}\label{b2}
{\mathsf W}_{dim-5}^{(10)}
=-2\lambda_{\acute{a}\acute{b},\acute{c}\acute{d}}^{^{(10)}}
<\Psi_{(+)\acute{a}}^*|B|\Upsilon_{(-)\acute{b}c_i}><\Psi_{(+)\acute{c}}^*|B|\Upsilon_{(-)\acute{d}\bar
c_{i}}>,
\end{eqnarray}
where
\begin{eqnarray}\label{b3}
\lambda_{\acute{a}\acute{b},\acute{c}\acute{d}}^{^{(10)}}=
j_{\acute{a}\acute{b}}^{^{(10)}}j_{\acute{c}\acute{d}}^{^{(10)}}k_{_{{\cal
S}}}^{^{(10)}}\left[\widetilde{{\cal M}}^{^{(10)}}\left\{{\cal
M}^{^{(54)}}\widetilde{{\cal M}}^{^{(10)}} -\bf{1}\right\}
\right]_{{\cal S}{\cal S}'}k_{_{{\cal S}'}}^{^{(10)}},\nonumber\\
\widetilde{{\cal M}}^{^{(10)}}=\left[{\cal
M}^{^{(54)}}+\left({\cal M}^{^{(10)}}\right)^{\bf {T}}\right]^{-1}.
\end{eqnarray}

Mass terms in the last equation are given by
\begin{eqnarray}\label{b4}
{\mathsf
W}^{^{(10)'}}_{mass}=\lambda_{\acute{a}\acute{b},\acute{c}\acute{d}}^{^{(10)}}\left[
 -\frac{1}{\sqrt 5}{\bf\widehat M}_{\acute{a}}^{ij\bf
T}{{\bf\widehat Q}}_{\acute{b}j}{\bf\widehat M}_{\acute{c}k}^{\bf T}{{\bf\widehat Q}}_{\acute{d}i}^{k}+2{\bf\widehat M}_{\acute{a}i}^{\bf T}{{\bf\widehat Q}}_{\acute{b}}^{j}{\bf\widehat M}_{\acute{c}}^{\bf T}{{\bf\widehat Q}}_{\acute{d}j}^{i}
+\frac{4}{\sqrt 5}{\bf\widehat M}_{\acute{a}}^{\bf T}{{\bf\widehat Q}}_{\acute{b}}^{i}{\bf\widehat M}_{\acute{c}}^{\bf T}{{\bf\widehat Q}}_{\acute{d}i}\right].
\end{eqnarray}

\subsection{$\mathbf{(16\cdot
\overline{144})_{45}(16 \cdot\overline{144})_{45}}$ coupling}
To generate  the $\mathbf{(16\cdot
\overline{144})_{45}(16 \cdot\overline{144})_{45}}$ coupling
we consider the following superpotential
 \begin{equation}\label{b5}
 {\mathsf W}^{^{(45)}}=\frac{1}{2}\Phi_{\mu\nu {\cal X}} {\cal
M}^{^{(45)}}_{{\cal X}{\cal X}'} \Phi_{\mu\nu {\cal
X}'}+\frac{1}{2!}h_{{\acute{a}\acute{b}}}^{^{(45)}}\Omega_{{\mu\nu,\acute{a}\acute{b}}}\Phi_{\mu\nu
{\cal X}}k_{_{{\cal X}}}^{^{(45)}},
 \end{equation}
 where $\Phi_{\mu\nu {\cal X}}$ is the $45$ plet field and
 \begin{equation}\label{b6}
\Omega_{{\mu\nu,\acute{a}\acute{b}}}=<\Psi^*_{(+)\acute{a}}|B\Gamma
_{\mu}|\Psi_{(+)\acute{b}\nu}>-<\Psi^*_{(+)\acute{a}}|B\Gamma
_{\nu}|\Psi_{(+)\acute{b}\mu}>.
\end{equation}
 After elimination of the heavy $45$ of $SO(10)$ we get,
\begin{eqnarray}\label{b7}
{\mathsf W}^{^{(16\cdot {\overline {144}})_{45}(16\cdot
{\overline
{144}})_{45}}}=-\frac{1}{2}\lambda_{\acute{a}\acute{b},\acute{c}\acute{d}}^{^{(45)}}\Omega_{{\mu\nu,\acute{a}\acute{b}}}
\Omega_{{\mu\nu,\acute{c}\acute{d}}}\nonumber\\
=\lambda_{\acute{a}\acute{b},\acute{c}\acute{d}}^{^{(45)}}\left[
-2<\Psi_{(+)\acute{a}}^*|Bb_i|\Psi_{(+)\acute{b}c_j}><\Psi_{(+)\acute{c}}^*|Bb_i^{\dagger}|\Psi_{(+)\acute{d}\bar
c_{j}}>\right.\nonumber\\
\left.+2<\Psi_{(+)\acute{a}}^*|Bb_i|\Psi_{(+)\acute{b}c_j}><\Psi_{(+)\acute{c}}^*|Bb_j^{\dagger}|\Psi_{(+)\acute{d}\bar
c_{i}}>\right.\nonumber\\
\left.-2<\Psi_{(+)\acute{a}}^*|Bb_i|\Psi_{(+)\acute{b}\bar
c_j}><\Psi_{(+)\acute{c}}^*|Bb_i^{\dagger}|\Psi_{(+)\acute{d}
c_{j}}>\right.\nonumber\\
\left.+<\Psi_{(+)\acute{a}}^*|Bb_i|\Psi_{(+)\acute{b}\bar
c_j}><\Psi_{(+)\acute{c}}^*|Bb_j|\Psi_{(+)\acute{d}\bar
c_{i}}>\right.\nonumber\\
\left.+<\Psi_{(+)\acute{a}}^*|Bb_i^{\dagger}|\Psi_{(+)\acute{b}
c_j}><\Psi_{(+)\acute{c}}^*|Bb_j^{\dagger}|\Psi_{(+)\acute{d}
c_{i}}>\right],
\end{eqnarray}
where we have defined
\begin{eqnarray}\label{b8}
\lambda_{\acute{a}\acute{b},\acute{c}\acute{d}}^{^{(45)}}=
h_{\acute{a}\acute{b}}^{^{(45)}}h_{\acute{c}\acute{d}}^{^{(45)}}k_{_{{\cal
X}}}^{^{(45)}}\left[\widetilde{{\cal M}}^{^{(45)}}\left\{{\cal
M}^{^{(45)}}\widetilde{{\cal M}}^{^{(45)}} -\bf{1}\right\}
\right]_{{\cal X}{\cal X}'}k_{_{{\cal X}'}}^{^{(45)}},\nonumber\\
\widetilde{{\cal M}}^{^{(45)}}=\left[{\cal
M}^{^{(45)}}+\left({\cal M}^{^{(45)}}\right)^{\bf {T}}\right]^{-1}.
\end{eqnarray}
From the above we find the following couplings which contribute to
the quark, lepton and neutrino masses\cite{ns}
\begin{eqnarray}\label{b9}
{\mathsf
W}^{^{(45)}}_{mass}=\lambda_{\acute{a}\acute{b},\acute{c}\acute{d}}^{^{(45)}}\left[
\epsilon_{ijklm}\left(-\frac{1}{\sqrt 5}{\bf\widehat M}_{\acute{a}}^{ix\bf T}{{\bf\widehat P}}_{\acute{b}x}^{j}{\bf\widehat M}_{\acute{c}}^{kl\bf T}{{\bf\widehat P}}_{\acute{d}}^{m}+\frac{1}{2}{\bf\widehat M}_{\acute{a}}^{xy\bf T}{{\bf\widehat P}}_{\acute{b}y}^{i}{\bf\widehat M}_{\acute{c}}^{jk\bf T}{{\bf\widehat P}}_{\acute{d}x}^{lm}-\frac{1}{2}{\bf\widehat M}_{\acute{a}}^{ix\bf
T}{{\bf\widehat P}}_{\acute{b}x}^{y}{\bf\widehat M}_{\acute{c}}^{jk\bf T}{{\bf\widehat P}}_{\acute{d}y}^{lm}\right)\right.\nonumber\\
\left.+2{\bf\widehat M}_{\acute{a}}^{ij\bf T}\left({{\bf\widehat P}}_{\acute{b}j}^{k}{\bf\widehat M}_{\acute{c}i}^{\bf T}{{\bf\widehat P}}_{\acute{d}k}-{{\bf\widehat P}}_{\acute{b}j}^{k}{\bf\widehat M}_{\acute{c}k}^{\bf T}{{\bf\widehat P}}_{\acute{d}i}\right)\right.
\left.+{\bf\widehat M}_{\acute{a}i}^{\bf T}\left(-\frac{3}{\sqrt 5}{{\bf\widehat P}}_{\acute{b}}^j {\bf\widehat M}_{\acute{c}}^{\bf T}{{\bf\widehat P}}_{\acute{d}j}^i-2{{\bf\widehat P}}_{\acute{b}k}^{ij}{\bf\widehat M}_{\acute{c}}^{\bf T}{{\bf\widehat P}}_{\acute{dj}}^k\right)\right.\nonumber\\
\left.-{\bf\widehat M}_{\acute{a}}^{\bf T}{{\bf\widehat P}}_{\acute{b}j}^i{\bf\widehat M}_{\acute{c}}^{\bf T}{{\bf\widehat P}}_{\acute{d}i}^j\right.
\left.+{\bf\widehat M}_{\acute{a}i}^{\bf T}\left(-\frac{1}{4}{{\bf\widehat P}}_{\acute{b}}^{i}{\bf\widehat M}_{\acute{c}j}^{\bf
T}{\widehat{\bf\widehat P}}_{\acute{d}}^{j}-\frac{5}{4}{{\bf\widehat P}}_{\acute{b}}^{j}{\bf\widehat M}_{\acute{c}j}^{\bf
T}{\widehat{\bf\widehat P}}_{\acute{d}}^{i}\right)\right].
\end{eqnarray}

\subsection{$\mathbf{(16\cdot
\overline{144})_{54}(16 \cdot\overline{144})_{54}}$ coupling}

 The 54-dimensional representation of $SO(10)$ is given by the tensor $\Phi_{\mu\nu}$ which is  symmetric and
 traceless and is given by
 \begin{equation}\label{b11}
\Delta_{{\mu\nu,\acute{a}\acute{b}}}=\Xi_{{\mu\nu,\acute{a}\acute{b}}}-\frac{1}{10}\delta_{\mu\nu}\Xi_{{\sigma\sigma,\acute{a}\acute{b}}},
\end{equation}
where
 \begin{equation}\label{b12}
\Xi_{{\mu\nu,\acute{a}\acute{b}}}=<\Psi^*_{(+)\acute{a}}|B\Gamma
_{\mu}|\Psi_{(+)\acute{b}\nu}>+<\Psi^*_{(+)\acute{a}}|B\Gamma
_{\nu}|\Psi_{(+)\acute{b}\mu}>.
\end{equation}
To generate the desired coupling we
 consider the
following superpotential
 \begin{equation}\label{b10}
 {\mathsf W}^{^{(54)}}=\frac{1}{2}\Phi_{\mu\nu {\cal T}} {\cal
M}^{^{(54)}}_{{\cal T}{\cal T}'} \Phi_{\mu\nu {\cal
T}'}+\frac{1}{2!}h_{{\acute{a}\acute{b}}}^{^{(54)}}\Delta_{{\mu\nu,\acute{a}\acute{b}}}\Phi_{\mu\nu
{\cal T}}k_{\cal T}^{^{(54)}}.
 \end{equation}
After elimination of the heavy $54$ of $SO(10)$ using the
F-flatness conditions: $\frac{\partial{\mathsf
W}^{(54)}}{\partial\Phi_{\mu\nu\cal{T}}}$ we get,
\begin{eqnarray}\label{b13}
{\mathsf
W}_{dim-5}^{(54)}=-\frac{1}{2}\lambda_{\acute{a}\acute{b},\acute{c}\acute{d}}^{^{(45)}}\Delta_{{\mu\nu,\acute{a}\acute{b}}}
\Delta_{{\mu\nu,\acute{c}\acute{d}}},
\end{eqnarray}
where we have defined
\begin{eqnarray}\label{b14}
\lambda_{\acute{a}\acute{b},\acute{c}\acute{d}}^{^{(54)}}=
h_{\acute{a}\acute{b}}^{^{(54)}}h_{\acute{c}\acute{d}}^{^{(54)}}k_{_{{\cal
T}}}^{^{(54)}}\left[\widetilde{{\cal M}}^{^{(54)}}\left\{{\cal
M}^{^{(54)}}\widetilde{{\cal M}}^{^{(54)}} -\bf{1}\right\}
\right]_{{\cal T}{\cal T}'}k_{_{{\cal T}'}}^{^{(54)}},\nonumber\\
\widetilde{{\cal M}}^{^{(54)}}=\left[{\cal
M}^{^{(54)}}+\left({\cal M}^{^{(54)}}\right)^{\bf {T}}\right]^{-1}.
\end{eqnarray}
Using Eqs.(\ref{b11}) and (\ref{b12}) to expand Eq.(\ref{b13}), we obtain
\begin{eqnarray}\label{b15}
{\mathsf W}_{dim-5}^{(54)}
=\lambda_{\acute{a}\acute{b},\acute{c}\acute{d}}^{^{(54)}}[
-2<\Psi_{(+)\acute{a}}^*|Bb_i|\Upsilon_{(+)\acute{b}c_j}><\Psi_{(+)\acute{c}}^*|Bb_i^{\dagger}|\Upsilon_{(+)\acute{d}\bar
c_{j}}>\nonumber\\
-2<\Psi_{(+)\acute{a}}^*|Bb_i|\Upsilon_{(+)\acute{b}c_j}><\Psi_{(+)\acute{c}}^*|Bb_j^{\dagger}|\Upsilon_{(+)\acute{d}\bar
c_{i}}>\nonumber\\
-2<\Psi_{(+)\acute{a}}^*|Bb_i|\Upsilon_{(+)\acute{b}\bar
c_j}><\Psi_{(+)\acute{c}}^*|Bb_i^{\dagger}|\Upsilon_{(+)\acute{d}
c_{j}}>\nonumber\\
-<\Psi_{(+)\acute{a}}^*|Bb_i|\Upsilon_{(+)\acute{b}\bar
c_j}><\Psi_{(+)\acute{c}}^*|Bb_j|\Upsilon_{(+)\acute{d}\bar
c_{i}}>\nonumber\\
-<\Psi_{(+)\acute{a}}^*|Bb_i^{\dagger}|\Upsilon_{(+)\acute{b}
c_j}><\Psi_{(+)\acute{c}}^*|Bb_j^{\dagger}|\Upsilon_{(+)\acute{d}
c_{i}}>\nonumber\\
+\frac{4}{5}<\Psi_{(+)\acute{a}}^*|Bb_i|\Upsilon_{(+)\acute{b}\bar
c_i}><\Psi_{(+)\acute{c}}^*|Bb_j|\Upsilon_{(+)\acute{d}\bar
c_{j}}>\nonumber\\
+\frac{4}{5}<\Psi_{(+)\acute{a}}^*|Bb_i^{\dagger}|\Upsilon_{(+)\acute{b}
c_i}><\Psi_{(+)\acute{c}}^*|Bb_j^{\dagger}|\Upsilon_{(+)\acute{d}
c_{j}}>\nonumber\\
+\frac{8}{5}<\Psi_{(+)\acute{a}}^*|Bb_i|\Upsilon_{(+)\acute{b}\bar
c_i}><\Psi_{(+)\acute{c}}^*|Bb_j^{\dagger}|\Upsilon_{(+)\acute{d}
c_{j}}>].
\end{eqnarray}

Finally, we look for the terms in the equation above which
contribute to the quark, lepton and neutrino masses
\begin{eqnarray}\label{b16}
{\mathsf
W}^{^{(54)}}_{mass}=\lambda_{\acute{a}\acute{b},\acute{c}\acute{d}}^{^{(54)}}\left[
 -\frac{1}{2}\epsilon_{ijklm}{\bf\widehat M}_{\acute{a}}^{xy\bf
T}{{\bf\widehat P}}_{\acute{b}y}^{k}{\bf\widehat M}_{\acute{c}}^{ij\bf T}{{\bf\widehat P}}_{\acute{d}x}^{lm}-\frac{1}{2}\epsilon_{ijklm}{\bf\widehat M}_{\acute{a}}^{ix\bf T}{{\bf\widehat P}}_{\acute{b}x}^{y}{\bf\widehat M}_{\acute{c}}^{jk\bf T}{{\bf\widehat P}}_{\acute{d}y}^{lm}\right.\nonumber\\
\left.-2{\bf\widehat M}_{\acute{a}}^{ij\bf T}{{\bf\widehat P}}_{\acute{b}i}^{k}{\bf\widehat M}_{\acute{c}j}^{\bf T}{{\bf\widehat P}}_{\acute{d}k}+2{\bf\widehat M}_{\acute{a}}^{ij\bf T}{{\bf\widehat P}}_{\acute{b}j}^{k}{\bf\widehat M}_{\acute{c}k}^{\bf T}{{\bf\widehat P}}_{\acute{d}i}
+\frac{1}{\sqrt 5 }{\bf\widehat M}_{\acute{a}i}^{\bf T}{{\bf\widehat P}}_{\acute{b}}^{j}{\bf\widehat M}_{\acute{c}}^{\bf T}{{\bf\widehat P}}_{\acute{d}j}^{i}-2{\bf\widehat M}_{\acute{a}i}^{\bf T}{{\bf\widehat P}}_{\acute{b}k}^{ij}{\bf\widehat M}_{\acute{c}}^{\bf T}{{\bf\widehat P}}_{\acute{d}j}^{k}\right.\nonumber\\
\left.+{\bf\widehat M}_{\acute{a}}^{\bf T}{{\bf\widehat P}}_{\acute{b}j}^{i}{\bf\widehat M}_{\acute{c}}^{\bf T}{{\bf\widehat P}}_{\acute{d}i}^{j}-\frac{201}{100}{\bf\widehat M}_{\acute{a}i}^{\bf
T}{{\bf\widehat P}}_{\acute{b}}^{i}{\bf\widehat M}_{\acute{c}j}^{\bf T}{{\bf\widehat P}}_{\acute{d}}^{j}\right]
+\frac{9}{20}\lambda_{\acute{a}\acute{d},\acute{c}\acute{b}}^{^{(54)}}{\bf\widehat M}_{\acute{a}i}^{\bf T}{{\bf\widehat P}}_{\acute{b}}^{i}{\bf\widehat M}_{\acute{c}j}^{\bf T}{{\bf\widehat P}}_{\acute{d}}^{j}.
\end{eqnarray}

\section{Appendix C: General formulae on quark masses from quartic couplings\label{c}}
We give here the general formulae for the quark and lepton masses including all the relevant
interactions. For the up quark we have

\begin{eqnarray}\label{c1}
M^{u}&=& \sum_{i=1}^{5}A^{(10)}_{i} + \sum_{i=1}^{4}A^{(45)}_{i}
+\sum_{i=1}^{2}A^{(54)}_{i}+ A^{(126)}.
\end{eqnarray}
For the down type quarks and for the charged  leptons we have
\begin{eqnarray}\label{c2}
 M^{d}_{\acute {a}\acute {b}}&\equiv &
\left({X} +{Y}\right)_{\acute {a}\acute{b}},~~~~~
M^{e}_{\acute {a}\acute {b}}\equiv\left({ X} -3{
Y}\right)_{\acute {a}\acute {b}},\label{c2}\\
\nonumber\\
X&=&\frac{1}{2} \sum_{i=1}^5 B^{(10)}_{i}
+
\frac{9}{8}(B^{(10)}_{6}+ B^{(10)}_{7})
+\frac{3}{4}(-B^{(45)}+B^{(54)})\nonumber\\
&&+\frac{21}{22}(B^{(126)}_{1} + B^{(126)}_{2})
+\frac{3}{2}(B^{(126)}_{3}+ B^{(126)}_{4}),\label{c3}\\
\nonumber\\
Y&=&\frac{1}{2} \sum_{i=1}^5 B^{(10)}_{i}
-\frac{1}{8}(B^{(10)}_{6}+B^{(10)}_{7})
+\frac{7}{4}B^{(45)}+\frac{1}{4}B^{(54)}\nonumber\\
&&+\frac{1}{22}(B^{(126)}_{1} + B^{(126)}_{2})
 -\frac{1}{2}(B^{(126)}_{3}+B^{(126)}_{4})\label{c4}.
\end{eqnarray}
For the Dirac neutrino we have
\begin{eqnarray}
M^{D\nu}&=&\sum_{i=1}^5 A^{(10)}_{i}
+\frac{9}{8}\left[1-\frac{5}{4}\frac{\lambda_{{\acute {a}}{\acute
{d}},{\acute {b}}{\acute {c}}}^{^{(45)}}}{\lambda_{{\acute
{a}}{\acute {c}},{\acute {b}}{\acute
{d}}}^{^{(45)}}}\right]^{-1}\left(A^{(45)}_{1}+A^{(45)}_{2}\right)
+\frac{15}{4}\left[1-\frac{9}{2}\frac{\lambda_{{\acute
{a}}{\acute {d}},{\acute {b}}{\acute
{c}}}^{^{(45)}}}{{\lambda_{{\acute {a}}{\acute {c}},{\acute
{b}}{\acute
{d}}}^{^{(45)}}}}\right]^{-1}\left(A^{(45)}_{3}+A^{(45)}_{4}\right)\nonumber\\
&&+\frac{15}{4}\left(A^{(54)}_{1}+A^{(54)}_{2}\right)+A^{(54)}_{3}+A^{(54)}_{4}
-3A^{(126)},\label{c5}\\
\nonumber\\
\nonumber\\
M^{RR}&=&C^{(45)}+C^{(54)}+C^{(10)}_{1}+C^{(10)}_{2}+C^{(126)}_{1}+C^{(126)}_{2},\label{c6}
\nonumber\\
\nonumber\\
M^{LL}&=&\sum_{i=1}^3(
D^{(45)}_i+D^{(54)}_{i})\label{c7}.
\end{eqnarray}
Subcases of these are  discussed in Sec.(3.1).

\section{Appendix D: Expansion of $\mathbf{(16\cdot 45\cdot \overline{144})}$ and $\mathbf{(16\cdot 120\cdot 144)}$ couplings
\label{d}}

We give in this Appendix an $SU(5)\times U(1)$ expansion of $\mathbf{(16\cdot 45\cdot \overline{144})}$ and $\mathbf{(16\cdot 120\cdot 144)}$ couplings. Thus for the $\mathbf{(16\cdot 45\cdot \overline{144})}$ coupling  we have the expansion
\begin{eqnarray}\label{d1}
{\mathsf W}^{16\cdot  \cdot {45}\cdot\overline{144}}&=&
2h_{\acute{a}\acute{b}}^{(45)}\left\{<{\widehat\Psi}_{(+)\acute{a}}^{*}|Bb_{i}|{\widehat\Upsilon}_{(+)c_j}>
\widehat{ {\bf
F}}_{\acute{b}ij}^{(45)}+<{\widehat\Psi}_{(+)\acute{a}}^{*}|Bb_{i}^{\dagger}|{\widehat\Upsilon}_{(+)\bar
c_j}> \widehat { {\bf F}}_{\acute{b}}^{(45)ij}\right.\nonumber\\
&&\left.+<{\widehat\Psi}_{(+)\acute{a}}^{*}|Bb_{i}^{\dagger}|{\widehat\Upsilon}_{(+)c_j}>
\widehat { {\bf
F}}_{\acute{b}j}^{(45)i}-<{\widehat\Psi}_{(+)\acute{a}}^{*}|Bb_{i}|{\widehat\Upsilon}_{(+)\bar
c_j}> \widehat { {\bf F}}_{\acute{b}i}^{(45)j}\right.\nonumber\\
&&
\left.+\frac{1}{5}\left[<{\widehat\Psi}_{(+)\acute{a}}^{*}|Bb_{n}^{\dagger}|{\widehat\Upsilon}_{(+)c_n}>-
 <{\widehat\Psi}_{(+)\acute{a}}^{*}|Bb_{n}|{\widehat\Upsilon}_{(+)\bar
 c_n}>\right]
 \widehat { {\bf F}}_{\acute{b}}^{(45)}
 \right\},\nonumber\\
 \nonumber\\
{\mathsf W}_{{{ mass}}}^{(45)}&=&m_F^{(45)}\left[\widehat { {\bf
F}}^{(45)ij}\widehat{ {\bf F}}_{ij}^{(45)}-\widehat { {\bf
F}}_{j}^{(45)i}\widehat { {\bf F}}_{i}^{(45)j}-\widehat { {\bf
F}}^{(45)}\widehat { {\bf F}}^{(45)}\right].
\end{eqnarray}
and for the $\mathbf{(16\cdot 120\cdot 144)}$ couplings we have the expansion

\begin{eqnarray}\label{d2}
{\mathsf W}^{16 \cdot{120}\cdot {144}}&=&h_{\acute{a}\acute{b}}^{(120)}\left\{
    4\epsilon_{ijklm}<{\widehat\Psi}_{(+)\acute{a}}^{*}|Bb_{i}b_{j}|{\widehat\Upsilon}_{(-)c_k}>
\widehat{ {\bf
F}}_{\acute{b}}^{(120)lm}\right.\nonumber\\
&+&\negthinspace\negthinspace\negthinspace\left.
4\epsilon^{ijklm}<{\widehat\Psi}_{(+)\acute{a}}^{*}|Bb_{i}^{\dagger}b_{j}^{\dagger}|{\widehat\Upsilon}_{(-)\bar
c_k}> \widehat{ {\bf F}}_{\acute{b}lm}^{(120)}
\right.\nonumber\\
&+&\negthinspace\negthinspace\negthinspace\left.\left[4<{\widehat\Psi}_{(+)\acute{a}}^{*}|Bb_{i}b_{j}|{\widehat\Upsilon}_{(-)\bar
c_k}>-8<{\widehat\Psi}_{(+)\acute{a}}^{*}|Bb_{k}^{\dagger}b_{j}|{\widehat\Upsilon}_{(-)c_i}>\right]\widehat{
{\bf F}}_{\acute{b}ij}^{(120)k} \right.\nonumber\\
&+&\negthinspace\negthinspace\negthinspace\left.
\left[4<{\widehat\Psi}_{(+)\acute{a}}^{*}|Bb_{i}^{\dagger}b_{j}^{\dagger}|{\widehat\Upsilon}_{(-)
c_k}>-8<{\widehat\Psi}_{(+)\acute{a}}^{*}|Bb_{i}^{\dagger}b_{k}|{\widehat\Upsilon}_{(-)\bar
c_j}>\right]\widehat{ {\bf F}}_{\acute{b}k}^{(120)ij}
\right.\nonumber\\
&+&\negthinspace\negthinspace\negthinspace\left.\left[-2<{\widehat\Psi}_{(+)\acute{a}}^{*}|Bb_{i}b_{n}|{\widehat\Upsilon}_{(-)
\bar
c_n}>-2<{\widehat\Psi}_{(+)\acute{a}}^{*}|Bb_{n}^{\dagger}b_{i}|{\widehat\Upsilon}_{(-)
 c_n}>+<{\widehat\Psi}_{(+)\acute{a}}^{*}|B\left(4+2b_{n}^{\dagger}b_{n}\right)|{\widehat\Upsilon}_{(-)
 c_i}>\right]\widehat{ {\bf
 F}}_{\acute{b}i}^{(120)}\right.\nonumber\\
&+&\negthinspace\negthinspace\negthinspace\left.\left[2<{\widehat\Psi}_{(+)\acute{a}}^{*}|Bb_{n}^{\dagger}b_{i}^{\dagger}|{\widehat\Upsilon}_{(-)
c_n}>+2<{\widehat\Psi}_{(+)\acute{a}}^{*}|Bb_{i}^{\dagger}b_{n}|{\widehat\Upsilon}_{(-)
 \bar c_n}>+<{\widehat\Psi}_{(+)\acute{a}}^{*}|B\left(4-2b_{n}^{\dagger}b_{n}\right)|{\widehat\Upsilon}_{(-)
 \bar c_i}>\right]\widehat{ {\bf
 F}}_{\acute{b}}^{(120)i}\right\},\nonumber\\
\nonumber\\
 {\mathsf W}_{{{
mass}}}^{(120)}&=&m_F^{(120)}\left[\widehat { {\bf
F}}^{(120)ij}\widehat{ {\bf F}}_{ij}^{(120)}+\widehat { {\bf
F}}_{ij}^{(120)k}\widehat { {\bf F}}_{k}^{(120)ij}+2\widehat {
{\bf F}}^{(120)}_{i}\widehat { {\bf F}}^{(120)i}\right].
\end{eqnarray}

\section{Appendix E: Contributions of ${\mathbf {16\cdot 45\cdot \overline{144}}}$ and
${\mathbf {16\cdot 120\cdot 144}}$ couplings to the ${\mathbf {b,\tau,t}}$ mass matrices\label{e}}
In this Appendix we give further details of the analysis of Sec.(\ref{4}).
Specifically we give here the terms that contribute to the mass matrices
for the $b$ quark, for the $\tau$ lepton, and for the top quark that arise from the
interactions of Eq.(\ref{1.8}). We  begin by listing the mass terms  for the $b$ quark.
We have
\begin{eqnarray}\label{e1}
{\mathsf L}_{2,{\textnormal {b}}}^{(45)}&=&-2\sqrt 2 f^{(45)}_{33}
p \left[{}^{({\overline{10}}_{45})}\!\overline{{\Large{\textnormal
b}}}_{{ R}\alpha}~ {}^{(10_{16})}\!{\Large {\textnormal b}}_{{
L}}^{\alpha}\right]+\textnormal{H.c.},\nonumber\\
{\mathsf L}_{4,{\textnormal
{b}}}^{(45)}&=&-2m_F^{(45)}\left[{}^{({\overline{10}}_{45})}\!\overline{{\Large
{\textnormal b}}}_{{ R}\alpha}~{}^{(10_{45})}\!{\Large
{\textnormal
b}}_{{ L}}^{\alpha} \right]+\textnormal{H.c.},\nonumber\\
{\mathsf L}_{5,{\textnormal {b}}}^{(45)}&=&-2\sqrt 2 f^{(45)}_{33}
<{\bf P}_5>
\left[{}^{({\overline{5}}_{16})}\!\overline{{\Large{\textnormal
b}}}_{{ R}\alpha}~ {}^{(10_{45})}\!{\Large {\textnormal b}}_{{
L}}^{\alpha}\right]+\textnormal{H.c.},\nonumber\\
{\mathsf L}_{3,{\textnormal {b}}}^{(120)}&=&\frac{16}{\sqrt 3}
f^{(120)}_{33} q
\left[{}^{({\overline{5}}_{16})}\!\overline{{\Large{\textnormal
b}}}_{{ R}\alpha}~ {}^{(5_{120})}\!{\Large {\textnormal b}}_{{
L}}^{\alpha}\right]+\textnormal{H.c.},\nonumber\\
{\mathsf L}_{4,{\textnormal {b}}}^{(120)}&=&-8\sqrt{\frac{3}{5}}
f^{(120)}_{33}<{\bf Q}_5>
\left[{}^{({\overline{5}}_{16})}\!\overline{{\Large{\textnormal
b}}}_{{ R}\alpha}~ {}^{(10_{120})}\!{\Large {\textnormal b}}_{{
L}}^{\alpha}\right]+\textnormal{H.c.},\nonumber\\
{\mathsf L}_{5,{\textnormal {b}}}^{(120)}&=&8\sqrt{\frac{3}{5}}
f^{(120)}_{33}<{\bf Q}_5>
\left[{}^{({\overline{5}}_{120})}\!\overline{{\Large{\textnormal
b}}}_{{ R}\alpha}~ {}^{(10_{16})}\!{\Large {\textnormal b}}_{{
L}}^{\alpha}\right]+\textnormal{H.c.},\nonumber\\
{\mathsf L}_{6,{\textnormal {b}}}^{(120)}&=&\frac{16}{3}
f^{(120)}_{33}<\widetilde{\bf Q}_5>
\left[{}^{({\overline{5}}_{120})}\!\overline{{\Large{\textnormal
b}}}_{{ R}\alpha}~ {}^{(10_{16})}\!{\Large {\textnormal b}}_{{
L}}^{\alpha}\right]+\textnormal{H.c.},\nonumber\\
{\mathsf L}_{8,{\textnormal {b}}}^{(120)}&=&-\frac{16}{\sqrt 3}
f^{(120)}_{33}q
\left[{}^{({\overline{10}}_{120})}\!\overline{{\Large{\textnormal
b}}}_{{ R}\alpha}~ {}^{(10_{16})}\!{\Large {\textnormal b}}_{{
L}}^{\alpha}\right]+\textnormal{H.c.},\nonumber\\
{\mathsf L}_{9,{\textnormal
{b}}}^{(120)}&=&-2m_F^{(120)}\left[{}^{({\overline{10}}_{120})}\!\overline{{\Large
{\textnormal b}}}_{{ R}\alpha}~{}^{(10_{120})}\!{\Large
{\textnormal
b}}_{{ L}}^{\alpha} \right]+\textnormal{H.c.},\nonumber\\
{\mathsf L}_{10,{\textnormal
{b}}}^{(120)}&=&-2m_F^{(120)}\left[{}^{({\overline{5}}_{120})}\!\overline{{\Large
{\textnormal b}}}_{{ R}\alpha}~{}^{(5_{120})}\!{\Large
{\textnormal b}}_{{ L}}^{\alpha} \right]+\textnormal{H.c.}.
\end{eqnarray}

Next,  we list the mass terms for the $\tau$ lepton. Here we have
\begin{eqnarray}\label{e2}
{\mathsf L}_{2,\mTau}^{(45)}&=&-12\sqrt 2 f^{(45)}_{33} p
\left[\!{}^{(10_{16})}
    \overline{ {\Tau}}_{{R}}~ {}^{({\overline{10}}_{45})}\!{\Tau}_{{
L}}\right]+\textnormal{H.c.},\nonumber\\
{\mathsf
L}_{4,\mTau}^{(45)}&=&-2m_F^{(45)}\left[{}^{(10_{45})}\!\overline{{\Tau}}_{{R}}~{}^{({\overline{10}}_{45})}\!{\Tau}_{{ L}}^{\alpha} \right]+\textnormal{H.c.},\nonumber\\
{\mathsf L}_{5,\mTau}^{(45)}&=&2\sqrt 2 f^{(45)}_{33} <{\bf P}_5>
\left[{}^{(10_{45})}\!
    \overline{ {\Tau}}_{{R}}~ {}^{({\overline{5}}_{16})}\!{\Tau}_
        {{L}}\right]+\textnormal{H.c.},\nonumber\\
{\mathsf L}_{3,\mTau}^{(120)}&=&-8\sqrt {3} f^{(120)}_{33} q
\left[{}^{(5_{120})}\!\overline{ {\Tau}}_{{R}}~
{}^{({\overline{5}}_{16})}\!{\Tau}_{{
L}}\right]+\textnormal{H.c.},\nonumber\\
{\mathsf L}_{4,\mTau}^{(120)}&=&8\sqrt{\frac{3}{5}}
f^{(120)}_{33}<{\bf Q}_5> \left[{}^{(10_{120})}\!
    \overline{ {\Tau}}_{{R}}~ {}^{({\overline{5}}_{16})}\!{\Tau}_{{
L}}\right]+\textnormal{H.c.},\nonumber\\
{\mathsf L}_{5,\mTau}^{(120)}&=&-8\sqrt{\frac{3}{5}}
f^{(120)}_{33}<{\bf Q}_5> \left[{}^{(10_{16})}\!\overline{
{\Tau}}_{{R}}~ {}^{({\overline{5}}_{120})}\!{\Tau}_{{
L}}\right]+\textnormal{H.c.},\nonumber\\
{\mathsf L}_{6,\mTau}^{(120)}&=&16 f^{(120)}_{33}<\widetilde{\bf
Q}_5> \left[{}^{(10_{16})}\!
    \overline{ {\Tau}}_{{R}}~ {}^{({\overline{5}}_{120})}\!{\Tau}_{{
L}}\right]+\textnormal{H.c.},\nonumber\\
{\mathsf L}_{8,\mTau}^{(120)}&=&-32\sqrt{3} f^{(120)}_{33}q
\left[{}^{(10_{16})}\!\overline{ {\Tau}}_{{R}}~
{}^{({\overline{10}}_{120})}\!{\Tau}_{{
L}}\right]+\textnormal{H.c.},\nonumber\\
{\mathsf
L}_{9,\mTau}^{(120)}&=&-2m_F^{(120)}\left[{}^{(10_{120})}\!
    \overline{{\Tau}}_{{R}}~{}^{({\overline{10}}_{120})}\!{\Tau}_{{ L}} \right]+\textnormal{H.c.},\nonumber\\
{\mathsf
L}_{10,\mTau}^{(120)}&=&-2m_F^{(120)}\left[{}^{(5_{120})}{}\!\overline{{\Tau}}_{{R}}~^{({\overline{5}}_{120})}\!{\Tau}_{{
L}} \right]+\textnormal{H.c.}.
\end{eqnarray}
Finally we list the mass terms for the top quark. Here we have\footnote{We note that
${\mathsf L}_{7,  {\textnormal {t}}}^{(120)}$ contains ${\bf Q}^5$ while for the case of
the $16\cdot 10\cdot144$ plet couplings ${\bf Q}^5$ was  absent from the list of mass terms for
the top quark (see Eq.(43) of \cite{Babu:2006rp}). It is the presence of this term that allows
for the mass generation for the top quark for the light ${\mathsf D_1}$ doublet case in the analysis of Eq.(\ref{e3}).}
\begin{eqnarray}\label{e3}
{\mathsf L}_{1,
    {\textnormal {t}}}^{(45)}=2\sqrt{\frac{2}{5}}f^{(45)}_{33} <{\bf
P}^5>\left[{}^{(10_{16})}\!\overline{{\Large {\textnormal t}}}_{{
R}\alpha} ~ {}^{(10_{45})}\!{\Large {\textnormal t}}_{{
L}}^{\alpha}~+~{}^{(10_{45})}\!\overline{{\Large {\textnormal
t}}}_{{ R}\alpha}~ {}^{(10_{16})}\!{\Large {\textnormal t}}_{{
L}}^{\alpha}\right]+\textnormal{H.c.},\nonumber\\
{\mathsf L}_{2,{\textnormal {t}}}^{(45)}=2\sqrt 2 f^{(45)}_{33} p
\left[4~ {}^{(10_{16})}\!\overline{{\Large {\textnormal t}}}_{{
R}\alpha}~ {}^{({\overline{10}}_{45})}\!{\Large {\textnormal
t}}_{{
L}}^{\alpha}~-~{}^{({\overline{10}}_{45})}\!\overline{{\Large
{\textnormal t}}}_{{ R}\alpha}~ {}^{(10_{16})}\!{\Large
{\textnormal t}}_{{
L}}^{\alpha}\right]+\textnormal{H.c.},\nonumber\\
{\mathsf L}_{3,{\textnormal
{t}}}^{(45)}=2\sqrt{\frac{2}{3}}f^{(45)}_{33} <\widetilde{\bf
P}^5>\left[{}^{(10_{16})}\!\overline{{\Large {\textnormal t}}}_{{
R}\alpha} ~ {}^{(10_{45})}\!{\Large {\textnormal t}}_{{
L}}^{\alpha}~+~{}^{(10_{45})}\!\overline{{\Large {\textnormal
t}}}_{{ R}\alpha}~ {}^{(10_{16})}\!{\Large {\textnormal t}}_{{
L}}^{\alpha}\right]+\textnormal{H.c.},\nonumber\\
{\mathsf L}_{4,{\textnormal
{t}}}^{(45)}=-2m_F^{(45)}\left[{}^{(10_{45})}\!\overline{{\Large
{\textnormal t}}}_{{
R}\alpha}~{}^{({\overline{10}}_{45})}\!{\Large {\textnormal t}}_{{
L}}^{\alpha}~+~{}^{({\overline{10}}_{45})}\!\overline{{\Large
{\textnormal t}}}_{{ R}\alpha}~{}^{(10_{45})}\!{\Large
{\textnormal
t}}_{{ L}}^{\alpha} \right]+\textnormal{H.c.},\nonumber\\
{\mathsf L}_{7,
    {\textnormal {t}}}^{(120)}=-\frac{8}{\sqrt 3}f^{(120)}_{33} <{\bf
Q}^5>\left[{}^{(10_{16})}\!\overline{{\Large {\textnormal t}}}_{{
R}\alpha} ~ {}^{(10_{120})}\!{\Large {\textnormal t}}_{{
L}}^{\alpha}~+~{}^{(10_{120})}\!\overline{{\Large {\textnormal
t}}}_{{ R}\alpha}~ {}^{(10_{16})}\!{\Large {\textnormal t}}_{{
L}}^{\alpha}\right]+\textnormal{H.c.},\nonumber\\
{\mathsf L}_{8,{\textnormal {t}}}^{(120)}=\frac{16}{\sqrt 3}
f^{(120)}_{33} q \left[4~ {}^{(10_{16})}\!\overline{{\Large
{\textnormal t}}}_{{ R}\alpha}~
{}^{({\overline{10}}_{120})}\!{\Large {\textnormal t}}_{{
L}}^{\alpha}~-~{}^{({\overline{10}}_{120})}\!\overline{{\Large
{\textnormal t}}}_{{ R}\alpha}~ {}^{(10_{16})}\!{\Large
{\textnormal t}}_{{
L}}^{\alpha}\right]+\textnormal{H.c.},\nonumber\\
{\mathsf L}_{9,{\textnormal
{t}}}^{(120)}=-2m_F^{(120)}\left[{}^{(10_{120})}\!\overline{{\Large
{\textnormal t}}}_{{
R}\alpha}~{}^{({\overline{10}}_{120})}\!
    {\Large {\textnormal t}}_{{
L}}^{\alpha}~+~{}^{({\overline{10}}_{120})}\!\overline{{\Large
{\textnormal t}}}_{{ R}\alpha}~{}^{(10_{120})}\!{\Large
{\textnormal t}}_{{ L}}^{\alpha} \right]+\textnormal{H.c.}.
\end{eqnarray}
The mass  matrices in Sec.(\ref{4}) are constructed using the results
given above.

\section{Appendix F:
Mixing angles in the diagonalization of the $b,\tau,t$ mass matrices\label{f}}
We define now various quantities that enter in the analysis of Sec.(\ref{4}).
Thus in Sec.(\ref{4}) the quantity $V_b$ involves $b_1,b_2,b_3, b_{4\pm}, b_{5\pm}$ which are
defined  as follows:
\begin{eqnarray}\label{f1}
b_1&=&\frac{1}{\sqrt{1+\frac{1}{y_{45}^2}+\frac{32}{3}\frac{1}{y_{120}^2}}},\nonumber\\
b_2&=&\frac{-1}{\sqrt{1+{y_{45}^2}+\frac{32}{3}\left(\frac{y_{45}}{y_{120}}\right)^2}},\nonumber\\
b_3&=&\frac{-1}{\sqrt{1+\frac{3}{32}{y_{120}^2}+\frac{3}{32}\left(\frac{y_{120}}{y_{45}}\right)^2}},\nonumber\\
b_{4\pm}&=&\frac{1}{2}\frac{m_D^{(45)^2}\left(1-y_{45}^2\right)+m_D^{(120)^2}\left(1+\frac{3}{32}y_{120}^2\right)
\pm \sqrt{
\beta_b}}{m_D^{(45)^2}y_{45}\left[1-\frac{3}{32}\left(\frac{m_D^{(120)}}{m_D^{(45)}}\frac{y_{120}}{y_{45}}
\right)^2\right]},\nonumber\\
b_{5\pm}&=&-2\sqrt{\frac{2}{3}}\frac{m_D^{(45)^2}\left(1+y_{45}^2\right)+m_D^{(120)^2}\left(1-\frac{3}{32}y_{120}^2\right)
\pm \sqrt{
\beta_b}}{m_D^{(120)^2}y_{120}\left[1-\frac{32}{3}\left(\frac{m_D^{(45)}}{m_D^{(120)}}\frac{y_{45}}{y_{120}}
\right)^2\right]}.
\end{eqnarray}
Further, the  heavy modes of the  $b$-quark are given by
\begin{eqnarray}\label{f2}
 &\lambda_{{
b}_2}^2\approx \Lambda_{{
b}_2}^2=m_D^{(120)^2}\left(1+\frac{3}{32}y_{120}^2\right),&\nonumber\\
&\lambda_{{ b}_{3,4}}^2 \approx \Lambda_{{
b}_{3,4}}^2=\frac{1}{2}\left(\alpha_b\pm\sqrt{\beta_b}\right),&
\end{eqnarray}
where $\alpha_b$ and $\beta_b$ are given by
\beqn\label{f3}
\alpha_b\equiv
m_D^{(45)^2}\left(1+y_{45}^2\right)+m_D^{(120)^2}\left(1+\frac{3}{32}y_{120}^2\right),\nonumber\\
\beta_b\equiv\alpha_b^2-4m_D^{(120)^2}\left(m_D^{(45)}y_{45}\right)^2\left(1+\frac{3}{64}y_{120}^2\right)
-\frac{3}{8}m_D^{(45)^2}\left(m_D^{(120)}y_{120}\right)^2\left(1+\frac{1}{2}y_{45}^2\right).
\eeqn
In Sec.(\ref{4})  $V_{\tau}$ contains the quantities  $\tau_1,\tau_2, \tau_3, \tau_{4\pm}, \tau_{5\pm}$
which are defined below
\begin{eqnarray}\label{f5}
\tau_1&=&\frac{1}{\sqrt{1+\frac{36}{y_{45}^2}+\frac{384}{y_{120}^2}}},\nonumber\\
\tau_2&=&\frac{-1}{\sqrt{1+\frac{1}{36}{y_{45}^2}+\frac{32}{3}\left(\frac{y_{45}}{y_{120}}\right)^2}},\nonumber\\
\tau_3&=&\frac{-1}{\sqrt{1+\frac{1}{384}{y_{120}^2}+\frac{3}{32}\left(\frac{y_{120}}{y_{45}}\right)^2}},\nonumber\\
\tau_{4\pm}&=&-3\frac{m_E^{(45)^2}\left(1-\frac{1}{36}y_{45}^2\right)+16m_E^{(120)^2}\left(1+\frac{1}{384}y_{120}^2\right)
\pm \sqrt{
\beta_{\tau}}}{m_E^{(45)^2}y_{45}\left[1-\frac{3}{2}\left(\frac{m_E^{(120)}}{m_E^{(45)}}\frac{y_{120}}{y_{45}}
\right)^2\right]},\nonumber\\
\tau_{5\pm}&=&-\frac{1}{2}\sqrt{\frac{3}{2}}\frac{m_E^{(45)^2}
\left(1+\frac{1}{36}y_{45}^2\right)+16m_E^{(120)^2}\left(1-\frac{1}{384}y_{120}^2\right)
\pm \sqrt{
\beta_{\tau}}}{m_E^{(120)^2}y_{120}\left[1-\frac{2}{3}\left(\frac{m_E^{(45)}}{m_E^{(120)}}\frac{y_{45}}{y_{120}}
\right)^2\right]}.
\end{eqnarray}
The heavy eigen modes of the $\tau$ lepton are given by
\begin{eqnarray}\label{f6}
 &\lambda_{{
\tau}_2}^2\approx \Lambda_{{
\tau}_2}^2=m_E^{(120)^2}\left(1+\frac{1}{24}y_{120}^2\right),&\nonumber\\
&\lambda_{{ \tau}_{3,4}}^2 \approx \Lambda_{{
\tau}_{3,4}}^2=\frac{1}{2}\left(\alpha_{\tau}\pm\sqrt{\beta_{\tau}}\right),&
\end{eqnarray}
where $\alpha_{\tau}$ and $\beta_{\tau}$ are given by
\begin{eqnarray}\label{f7}
&\alpha_{\tau}\equiv
m_E^{(45)^2}\left(1+\frac{1}{36}y_{45}^2\right)+16m_E^{(120)^2}\left(1+\frac{1}{384}y_{120}^2\right),&\nonumber\\
&\beta_{\tau}\equiv\alpha_{\tau}^2-\frac{16}{9}m_E^{(120)^2}\left(m_E^{(45)}y_{45}\right)^2\left(1+\frac{1}{768}y_{120}^2\right)
-\frac{1}{6}m_E^{(45)^2}\left(m_E^{(120)}y_{120}\right)^2\left(1+\frac{1}{72}y_{45}^2\right).
&
\end{eqnarray}
Further,
 in Sec.(\ref{4}) the matrices $U_t, V_t$ contain the quantities $t_1,t_2,t_{3\pm}, t_{4\pm},
t_{5\pm}$ and $t'_1, t'_2, t'_{3\pm}, t'_{4\pm}, t'_{5\pm}$. They are defined by
\begin{eqnarray}\label{f8}
t_1&=&\frac{1}{4}\sqrt{\frac{3}{2}}y_{120},\nonumber\\
t_2&=&-\frac{1}{4}\sqrt{\frac{3}{2}}\frac{y_{120}}{y_{45}},\nonumber\\
t_{3\pm}&=&-2\sqrt{\frac{2}{3}}\frac{1}{m_U^{(120)^2}y_{120}}\left[m_U^{(45)^2}\left(1+y_{45}^2\right)
+m_U^{(120)^2}\left(1-\frac{1}{32}y_{120}^2\right) \pm \sqrt{
\beta_{t}}\right],\nonumber\\
t_{4\pm}&=&-2\sqrt{\frac{2}{3}}\frac{y_{45}}{y_{120}}\frac{1}{m_U^{(120)^2}}\left[m_U^{(45)^2}\left(1+y_{45}^2\right)
-m_U^{(120)^2}\left(1+\frac{1}{32}y_{120}^2\right) \pm \sqrt{
\beta_{t}}\right],\nonumber\\
t_{5\pm}&=&\frac{1}{32m_U^{(45)}m_U^{(120)}}\left[16m_U^{(45)^2}\left(1+\frac{1}{16}y_{45}^2\right)
-16m_U^{(120)^2}\left(1+\frac{3}{512}y_{120}^2\right) \pm \sqrt{
\beta_{t}^{\prime}}\right],
\end{eqnarray}
and the primed ones are given by
\begin{eqnarray}\label{f9}
t_1^{\prime}&=&\frac{1}{16}\sqrt{\frac{3}{2}}y_{120},\nonumber\\
t_2^{\prime}&=&\frac{1}{4}\sqrt{\frac{3}{2}}\frac{y_{120}}{y_{45}},\nonumber\\
t_{3\pm}^{\prime}&=&\frac{1}{2m_U^{(45)}m_U^{(120)}}\left[-m_U^{(45)^2}\left(1+y_{45}^2\right)
+m_U^{(120)^2}\left(1+\frac{1}{32}y_{120}^2\right) \pm \sqrt{
\beta_{t}}\right],\nonumber\\
t_{4\pm}^{\prime}&=&\frac{1}{\sqrt
6}\frac{1}{m_U^{(120)^2}y_{120}}\left[-16m_U^{(45)^2}\left(1+\frac{1}{16}y_{45}^2\right)
-16m_U^{(120)^2}\left(1-\frac{3}{512}y_{120}^2\right) \pm \sqrt{
\beta_{t}^{\prime}}\right],\nonumber\\
t_{5\pm}^{\prime}&=&\frac{1}{4\sqrt
6}\frac{y_{45}}{y_{120}}\frac{1}{m_U^{(120)^2}}\left[16m_U^{(45)^2}\left(1+\frac{1}{16}y_{45}^2\right)
-16m_U^{(120)^2}\left(1+\frac{3}{512}y_{120}^2\right) \pm \sqrt{
\beta_{t}^{\prime}}\right].
\end{eqnarray}
The heavy eigen modes of the top quark are given by
\begin{eqnarray}\label{f10}
&\lambda_{{ t}_{2,3}}^2 \approx \Lambda_{{
t}_{2,3}}^2=\frac{1}{2}\left(\alpha_t\pm\sqrt{\beta_t}\right),&\nonumber\\
&\lambda_{{ t}_{4,5}}^2 \approx \Lambda_{{
t}_{4,5}}^2=\frac{1}{2}\left(\alpha_t^{\prime}\pm\sqrt{\beta_t^{\prime}}\right),
\end{eqnarray}
where $\alpha_t,   \beta_t, \alpha_t', \beta_t'$ are given by
\begin{eqnarray}\label{f11}
&\alpha_t\equiv
m_U^{(45)^2}\left(1+y_{45}^2\right)+m_U^{(120)^2}\left(1+\frac{3}{32}y_{120}^2\right),&\nonumber\\
&\alpha_t^{\prime}\equiv
16m_U^{(45)^2}\left(1+\frac{1}{16}y_{45}^2\right)+16m_U^{(120)^2}\left(1+\frac{3}{512}y_{120}^2\right),&\nonumber\\
&\beta_t\equiv\alpha_t^2-4m_U^{(120)^2}\left(m_U^{(45)}y_{45}\right)^2\left(1+\frac{3}{64}y_{120}^2\right),
-\frac{3}{8}m_U^{(45)^2}\left(m_U^{(120)}y_{120}\right)^2\left(1+\frac{1}{2}y_{45}^2\right),
&\nonumber\\
&\beta_t^{\prime}\equiv\alpha_t^{\prime
2}-64m_U^{(120)^2}\left(m_U^{(45)}y_{45}\right)^2\left(1+\frac{3}{1024}y_{120}^2\right)
-6m_U^{(45)^2}\left(m_U^{(120)}y_{120}\right)^2\left(1+\frac{1}{32}y_{45}^2\right).
&
\end{eqnarray}

\section{Appendix G: Yukawa coupling parameters $\epsilon_{b,\tau,t}$ and $\delta_{b,\tau,t}$\label{g}}
The quantities $\epsilon_{b}$ and $\delta_{b}$ that enter in Sec.(\ref{4}) are defined by
\beqn\label{g1}
\epsilon_b=
\frac{9}{128}\frac{\left(m_D^{(45)}m_D^{(120)}\right)^2
y_{120}^4}{\left(1+\frac{3}{32}y_{120}^2\right)\Lambda_{{
b}_3}^2\Lambda_{{ b}_4}^2},
~~\delta_b=
\frac{8}{15}\frac{\left(m_D^{(45)}y_{45}\right)^2\left(m_D^{(120)}y_{120}\right)^2
y_{120}^4}{\left(1+\frac{3}{32}y_{120}^2\right)\Lambda_{{
b}_3}^2\Lambda_{{ b}_4}^2},
\eeqn
where
$\Lambda_{{ b}_3}$ and  $\Lambda_{{ b}_4}$ are defined by Eq.(\ref{f2}). The $y-$quantities are defined through
\begin{eqnarray}\label{g2}
&y_{45}\equiv \frac{m_F^{(45)}}{\sqrt 2
f^{(45)}_{33}p},~~~y_{120}\equiv \frac{m_F^{(120)}}{\sqrt 2
f^{(120)}_{33}q}.
\end{eqnarray}

Similarly the parameters $\epsilon_{\tau}$ and $\delta_{\tau}$  are defined by
\beqn\label{g3}
\epsilon_{\tau}=
\frac{1}{72}\frac{\left(m_E^{(45)}m_E^{(120)}\right)^2
y_{120}^4}{\left(1+\frac{1}{24}y_{120}^2\right)\Lambda_{{
\tau}_3}^2\Lambda_{{ \tau}_4}^2},
~~\delta_{\tau}=
\frac{1}{135}\frac{\left(m_E^{(45)}y_{45}\right)^2\left(m_E^{(120)}y_{120}\right)^2
y_{120}^4}{\left(1+\frac{1}{24}y_{120}^2\right)\Lambda_{{
\tau}_3}^2\Lambda_{{\tau}_4}^2},
\eeqn
where
$\Lambda_{{ \tau}_3}$ and  $\Lambda_{{ \tau}_4}$ are defined by Eq.(\ref{f6}).
Finally,  the parameters $\epsilon_{t}$ and $\delta_{t}$ are given by

\beqn\label{g4}
\epsilon_t=
50\frac{\left(m_U^{(45)}m_U^{(120)}\right)^4
\left(y_{45}^2y_{120}\right)^2}{\Lambda_{{ t}_2}^2\Lambda_{{
t}_3}^2\Lambda_{{ t}_4}^2\Lambda_{{
t}_5}^2},
~~\delta_t=
\frac{27}{640}\frac{\left(m_U^{(45)}m_U^{(120)}\right)^4\left(y_{45}y_{120}^2\right)^2
}{\Lambda_{{ t}_2}^2\Lambda_{{ t}_3}^2\Lambda_{{ t}_4}^2\Lambda_{{
t}_5}^2},
\eeqn
where
$\Lambda_{{ t}_2-{ t}_5}$ are defined by Eq.(\ref{f10}).

\end{document}